\documentstyle[12pt]{article}
\def\PRL #1 #2 #3{{\sl Phys. Rev. Lett.} {\bf#1} (#2) #3}
\def\NPB #1 #2 #3{{\sl Nucl. Phys.} {\bf B#1} (#2) #3}
\def\NPBFS #1 #2 #3 #4{{\sl Nucl. Phys.} {\bf B#2} [FS#1] (#3) #4}
\def\CMP #1 #2 #3{{\sl Commun. Math. Phys.} {\bf #1} (#2) #3}
\def\PRD #1 #2 #3{{\sl Phys. Rev.} {\bf D#1} (#2) #3}
\def\PLA #1 #2 #3{{\sl Phys. Lett.} {\bf #1A} (#2) #3}
\def\PLB #1 #2 #3{{\sl Phys. Lett.} {\bf #1B} (#2) #3}
\def\JMP #1 #2 #3{{\sl J. Math. Phys.} {\bf #1} (#2) #3}
\def\PTP #1 #2 #3{{\sl Prog. Theor. Phys.} {\bf #1} (#2) #3}
\def\SPTP #1 #2 #3{{\sl Suppl. Prog. Theor. Phys.} {\bf #1} (#2) #3}
\def\AoP #1 #2 #3{{\sl Ann. of Phys.} {\bf #1} (#2) #3}
\def\PNAS #1 #2 #3{{\sl Proc. Natl. Acad. Sci. USA} {\bf #1} (#2) #3}
\def\RMP #1 #2 #3{{\sl Rev. Mod. Phys.} {\bf #1} (#2) #3}
\def\PR #1 #2 #3{{\sl Phys. Reports} {\bf #1} (#2) #3}
\def\AoM #1 #2 #3{{\sl Ann. of Math.} {\bf #1} (#2) #3}
\def\UMN #1 #2 #3{{\sl Usp. Mat. Nauk} {\bf #1} (#2) #3}
\def\FAP #1 #2 #3{{\sl Funkt. Anal. Prilozheniya} {\bf #1} (#2) #3}
\def\FAaIA #1 #2 #3{{\sl Functional Analysis and Its Application} {\bf
#1} (#2) #3}
\def\BAMS #1 #2 #3{{\sl Bull. Am. Math. Soc.} {\bf #1} (#2)
#3} \def\TAMS #1 #2 #3{{\sl Trans. Am. Math. Soc.} {\bf #1} (#2) #3}
\def\InvM #1 #2 #3{{\sl Invent. Math.} {\bf #1} (#2) #3}
\def\LMP #1 #2 #3{{\sl Letters in Math. Phys.} {\bf #1} (#2) #3}
\def\IJMPA #1 #2 #3{{\sl Int. J. Mod. Phys.} {\bf A#1} (#2) #3}
\def\AdM #1 #2 #3{{\sl Advances in Math.} {\bf #1} (#2) #3}
\def\RMaP #1 #2 #3{{\sl Reports on Math. Phys.} {\bf #1} (#2) #3}
\def\IJM #1 #2 #3{{\sl Ill. J. Math.} {\bf #1} (#2) #3}
\def\APP #1 #2 #3{{\sl Acta Phys. Polon.} {\bf #1} (#2) #3}
\def\TMP #1 #2 #3{{\sl Theor. Mat. Phys.} {\bf #1} (#2) #3}
\def\JPA #1 #2 #3{{\sl J. Physics} {\bf A#1} (#2) #3}
\def\JSM #1 #2 #3{{\sl J. Soviet Math.} {\bf #1} (#2) #3}
\def\MPLA #1 #2 #3{{\sl Mod. Phys. Lett.} {\bf A#1} (#2) #3}
\def\JETP #1 #2 #3{{\sl Sov. Phys. JETP} {\bf #1} (#2) #3}
\def\JETPL #1 #2 #3{{\sl  Sov. Phys. JETP Lett.} {\bf #1} (#2) #3}
\def\PHSA #1 #2 #3{{\sl Physica} {\bf A#1} (#2) #3}
\def\CQG #1 #2 #3{{\sl Class. Quantum Grav.} {\bf #1} (#2) #3}
\def\SJNP #1 #2 #3{{\sl Sov. J. Nucl. Phys. (Yadern.Fiz.)} {\bf #1} (#2) #3}
\def\a{\alpha}\def\b{\beta}\def\g{\gamma}\def\d{\delta}
\def\k{\kappa}
\def\Th{\Theta}\def\th{\theta}\def\Om{\Omega}\def\G{\Gamma}
\def\m{\mu}
\newcommand{\p}[1]{(\ref{#1})}
\begin{document}
\renewcommand{\thefootnote}{\arabic{footnote}}

%{\small
\begin{center}
{\large \bf  Generalized action principle and\\
extrinsic geometry for\\
N=1 superparticle.
}
\vspace{1cm}
%\footnote{Work supported in part by the
%International Science Foundation under the Grant RY No 9000,
%and by Ukrainian State Committee on Science and Technology under the
%contract No 2.3/664.
%}
\\ Igor A. Bandos

\vspace{0.5cm}
{\it Kharkov Institute of Physics and Technology}\\
{\it 310108, Kharkov, Ukraine}

\vspace{0.5cm}
and
\renewcommand{\thefootnote}{\dagger}
\\ Alexei Yu. Nurmagambetov
\footnote{On leave from the Kharkov Institute of Physics and Technology,
310108, Kharkov, Ukraine.}

\vspace{0.5cm}
{\it International Centre for Theoretical Physics}\\
{\it 34014, Trieste, Italy}\\

\vspace{0.5cm}
e-mail:  kfti@rocket.kharkov.ua\\

\vspace{1.5cm}
{\bf Abstract}
\end{center}

It is proposed the generalized action functional for N=1 superparticle in
D=3,4,6 and 10 space -- time dimensions. The superfield geometric approach
equations describing superparticle motion in terms of extrinsic geometry
of the worldline superspace are obtained on the base of the generalized
action.
The off -- shell superdiffeomorphism invariance (in the
rheonomic sense) of the superparticle generalized action is proved. It was
demonstrated that the half of the fermionic and one bosonic (super)fields
disappear from the generalized action in the analytical basis.

Superparticle interaction with Abelian gauge theory is considered in the
framework of this formulation.
The geometric approach equations describing superparticle motion in
Abelian background are obtained.

\vspace{0.8cm}

\newpage
\section{Introduction.}

Recently the generalized action functional for superstrings and super -- p
-- branes was proposed \cite{bsv}. It solved some problems of the standard
twistor -- like (doubly supersymmetric or world -- volume superfield)
formulations \cite{stv} -- \cite{bnsv} (see also Refs. in
\cite{bpstv}) related to the presence in the latter ones of the Lagrange
multiplier superfields, which lead to the appearance of additional
(undesirable) propagating degrees of freedom \cite{gs2,bpstv} and of the
infinitely reducible symmetries \cite{gs92,bnsv} in the formulations of
type IIB D=10 superstrings and D=11 supermembranes.

So, there is a hope that the generalized action principle can help to solve
the problem of completely covariant quantization of D=10 superstring
\footnote{Recently {\sl Lorentz} ($SO(1,3)$) covariant quantization of
heterotic string compactified on Calabi -- Yau manifold has been carried
out in \cite{berk} using conformal field theory approach. Most successful
previous attempts to reach a progress in Lorentz ($SO(1,D-1)$) covariant
quantization of D=10 superstring are ones presented in Refs.
\cite{nissimov,kallosh}.  All principal problems on this way were solved
in the frame of the so -- called twistor~--~like Lorentz~--~harmonic
formulation \cite{bzst} (see also \cite{bhbz0}, \cite{bzm}, \cite{bzp}),
however the calculation problems \cite{bzst} hamper a progress on this
line.  Partially covariant quantization was studied in
\cite{kallosh87,morozov88,kallosh95,carlip}. For recent attempts to find
alternative way see \cite{derizl}}.

At the same time the generalized action can be used to reproduce naturally
\cite{bandzer,bsv96} the doubly supersymmetric generalization \cite{bpstv}
of the geometric approach \cite{geo}, which can be useful for the more
deep understanding of the group -- theoretical and geometrical properties
of (super)string and (super)membrane theory
\cite{bandzer} -- \cite{viswan} and, in particular, for investigation of
the problems of coupling supersymmetric extended objects with natural
(supergravity) background.

Up to now the generalized action principle together with doubly
supersymmetric geometrical approach was not developed for the case of
superparticles. However, they can provide, in particular,
toy models for looking for a way of covariant quantization
in the framework of the generalized action principle approach
\footnote{The study of the covariant quantization problem is
not so simple subject due to non -- linear realization of some
symmetries in the considered formulation. This is the main
obstacle for straightforward application of known modern quantization
schemes and incites in a search of new versions of covariant quantization
to overcome this problem.}
and for studying the general properties of geometrical approach to
supersymmetric objects in flat and curved space -- time.
Moreover, the case of (super)particle has some peculiar features, which are
absent for superstring and super -- p -- branes, but have to be in
tensionless supersymmetric extended objects: the so -- called null --
superstrings and null -- super -- p -- branes \cite{z0,bhbz0}. These
objects become very attractive in relation to the string cosmology
problems \cite{zhelt,sanchez}. Besides, they may be related to the M --
theory compactifications to six dimensions \cite{39prime}.

So, this paper can be regarded as ABC of the generalized action
principle and geometric approach to supersymmetric objects (interacting
with background fields).
We construct the generalized action functional for N=1
superparticle in D=3,4,6,10, investigate its properties and symmetries,
derive equations of motion in terms of worldline superfield and
construct the geometric approach for superparticle.

On shell, the geometric approach equations describe free superparticle
motion and give not any new information. However, as it was demonstrated
in our paper, when we omit proper dynamical equation from the
consideration, the rest of the geometric approach equations are
appropriate for description of superparticle motion in any (Abelian) gauge
field.

As preliminary steps for looking for a way for covariant quantization in
the frame of the generalized action approach, we demonstrate that the
generalized action for superparticle satisfies the conditions of the off
-- shell superdiffeomorphism invariance (in rheonomic sense \cite{rheo})
and prove that in the analytical coordinate basis Sokatchev's effect of
disappearance of one bosonic and half of the fermionic (super)fields
(being the compensators for nonlinearly realized reparametrization and
worldline supersymmetry) \cite{sok,sok87} takes place in the generalized
action functional.

An investigation of superparticle interaction with Abelian gauge
(super)field in the generalized action principle approach demonstrates
that coupling leads to the restriction of background field strength by the
standard set of superspace constraints (compare with \cite{shapiro} and
Refs. therein).  We derive the geometric approach equations for N=1
superparticle interacting with Abelian gauge background and find the
correspondence of the superfields describing the extrinsic geometry of the
worldline superspace with worldline image of the gauge field strength
superfield.

{\underline{Notations}}: the underlined indices belong to a
target superspace, $(++,--)$ and $(+,-)$ denote $SO(1,1)$
weights, $(q,\dot{q},i)$ are the indices of spinor and vector
representations of $SO(D-2)$ (for D=10 $q$ and $\dot q$ are the {\bf s}
and {\bf c} spinor indices of $SO(8)$).

\section{N=1 superparticle in diverse dimensions.}

\subsection{The generalized action for superparticle and moving frame
variables.}

The generalized action describing superparticle in D=3,4,6,10 dimensions
has the following form
\begin{eqnarray}\label{c1}
S=\int_{{\cal{M}}^1}\,\,\rho^{++}E^{--}\equiv \int_{{\cal{M}}^1}\,\,\rho^{++}
(dX^{\underline{m}}-id\Th\G^{\underline{m}}\Th)u^{--}_{\underline{m}}
\end{eqnarray}
The Lagrangian one -- form
$$
{\cal L}=\rho^{++}E^{--}
$$
is integrated over one -- dimensional bosonic submanifold
$${\cal{M}}^1=\{(\tau,\eta^{q}(\tau))\}$$ of worldline superspace
$$\Sigma^{(1\vert D-2)}=\{(\tau,\eta^{q})\}$$ parametrized by the
proper~--~time $\tau$ and its Grassmann superpartner $\eta^{q}$
$(q=1,\dots,D~-~2)$. Thus, the coordinates of a target superspace of the
model, including D~--~dimensional space~--~time coordinates
$X^{\underline{m}}$ ($m=0,1,\dots,D-1$), their superpartners
$\Th^{\underline{\mu}}$ (${\underline{\mu}}=1,\dots,2(D-2)$)
as well as scalar density
$\rho^{++}$ and light -- like moving frame vector $u^{--}_{\underline{m}}$
(see below) shall be regarded in \p{c1} as worldline superfields
$$
X^{\underline{m}}=X^{\underline{m}}(\tau,\eta^{q}),\ \ \
\Th^{\underline{\mu}}=\Th^{\underline{\mu}}(\tau,\eta^{q}),\ \ \
\rho^{++}=\rho^{++}(\tau,\eta^{q}),\ \ \
u^{--}_{\underline{m}}=u^{--}_{\underline{m}}(\tau,\eta^{q})
$$
but taken on (an arbitrary) bosonic line ${\cal M}^1$
($\eta^{q}=\eta^{q}(\tau)$):
$$
X^{\underline{m}}=X^{\underline{m}}(\tau,\eta^{q}(\tau)),\ \ \
\Th^{\underline{\mu}}=\Th^{\underline{\mu}}(\tau,\eta^{q}(\tau)),\ \ \
\rho^{++}=\rho^{++}(\tau,\eta^{q}(\tau)),\ \ \
u^{--}_{\underline{m}}=u^{--}_{\underline{m}}(\tau,\eta^{q}(\tau))
$$
The number of $\eta^{q}$ is considered to be half of the number of spinor
target superspace variables $\Th^{\underline{\mu}}$, whose number coincides
with the dimension of the minimal spinor representation, being $2(D-2)$
(${\underline{\mu}}=1,\dots,2(D-2)$) for the considered cases D=3,4,6,10.

In distinction to the superstring and supermembrane cases \cite{bsv}, none
of the components of the worldline supereinbein superfields
$e^{A}_{M}(\tau,\eta^{q})$
$$
e^{A}{\equiv (e^{++},e^{+p})}=d\tau e^{A}_{\tau}
(\tau,\eta^{q})+d\eta^{q}e^{A}_{q}(\tau,\eta^{q})
$$
are involved into the action functional \p{c1}. However, they appear
in the decomposition of the external differential
$$d=e^{++}D_{++}+e^{+q}D_{+q},$$ where $D_{+q}$ and $D_{++}$ are the
covariant derivatives for worldline scalar superfields.

In \p{c1} $\rho^{++}$ is an arbitrary scalar density superfield and
$$
E^{--}=\Pi^{\underline{m}}u^{--}_{\underline{m}}=(dX^{\underline{m}}-
id\Th\G^{\underline{m}}\Th)u^{--}_{\underline{m}}
$$
is an appropriate part of arbitrary local (co) -- frame
$E^{\underline{A}}=(E^{\underline{a}},E^{\underline{\a}})$
of flat target superspace. Supervielbein $E^{\underline{A}}$ is related to
the basic supercovariant forms \cite{volkov}
\begin{equation}\label{b2}
\Pi^{\underline{m}}=dX^{\underline{m}}-id\Th\G^{\underline{m}}\Th,\qquad
d\Th^{\underline{\mu}}
\end{equation}
by $SO(1,D-1)$ rotations
$$
E^{\underline{A}}\equiv (E^{\underline{a}},E^{\underline{\a}})=
(\Pi^{\underline{m}},d\Th^{\underline{\mu}})
\left (
\matrix{u^{\underline{a}}_{\underline{m}} & 0 \cr
0 & v^{\underline{\a}}_{\underline{\mu}} \cr}
\right )
$$
$$
E^{\underline{a}}\equiv (E^{++},E^{--},E^{i})=
\Pi^{\underline{m}}u^{\underline{a}}_{\underline{m}}\equiv (
\Pi^{\underline{m}}u^{++}_{\underline{m}},
\Pi^{\underline{m}}u^{--}_{\underline{m}},
\Pi^{\underline{m}}u^{i}_{\underline{m}})
$$
\begin{equation}\label{rot}
E^{\underline{\a}}\equiv (E^{+q},E^{-\dot q})=
d\Th^{\underline{\mu}}v^{\underline{\a}}_{\underline{\mu}}\equiv (
d\Th^{\underline{\mu}}v^{+}_{\underline{\mu} q},
d\Th^{\underline{\mu}}v^{-}_{\underline{\mu}\dot q})
\end{equation}
which vector and spinor representations are given by matrices
$u^{\underline{a}}_{\underline{m}}=(u^{++}_{\underline{m}},
u^{--}_{\underline{m}},u^{i}_{\underline{m}})$ and
$v^{\underline{\a}}_{\underline{\mu}}=(v^{+}_{\underline{\mu} q},
v^{-}_{\underline{\mu}\dot q})$ respectively
\begin{equation}\label{a}
u^{\underline{a}}_{\underline{m}}=(u^{\pm\pm}_{\underline{m}},
u^{i}_{\underline{m}})\in SO(1,D-1)\Longleftrightarrow
u^{\underline{a}}_{\underline{m}}u^{\underline{m}\underline{b}}=
\eta^{\underline{ab}}
\end{equation}
\begin{equation}\label{aa}
v^{\underline{\a}}_{\underline{\mu}}=(v^{+}_{\underline{\mu} q},
v^{-}_{\underline{\mu}\dot q})\in Spin(1,D-1)
\end{equation}

The components $u^{\pm\pm}_{\underline{m}},u^{i}_{\underline{m}}$,
$v^{+}_{\underline{\mu} q},v^{-}_{\underline{\mu}\dot q}$ of these
matrices can be treated as the vector
and spinor Lorentz harmonic variables \cite{sok,sok87}, \cite{bhbz0} --
\cite{bzp}. Due to the $SO(1,1)\otimes{SO(D-2)}$ gauge symmetry of the
considered action \p{c1} (see below) they can be regarded
as coordinates of non -- compact coset space
$SO(1,D-1)/SO(1,1)\otimes{SO(D-2)}$ (see Refs.
\cite{sok,sok87}, \cite{bzst}--\cite{bzp}, \cite{bpstv} for details).

Specific point of superparticle theory is the presence of $K_{D-2}$ gauge
symmetry in the action \p{c1} (see below). As a result, regarding the
gauge symmetry $SO(1,1)\otimes SO(D-2)\otimes K_{D-2}$ as an
identification relation on the space of harmonic (moving frame)
variables, one can identifies them with coordinates of the compact coset
space $S^{D-2}=SO(1,D-1)/(SO(1,1)\otimes SO(D-2)\otimes K_{D-2})$
\cite{ghs}, which can be regarded as a celestial sphere \cite{rindler}.

Being the vector and spinor representation of the same $SO(1,D-1)$
rotation, $u$ and $v$ variables are related by the condition of the
conservation of D -- dimensional gamma -- matrices
\begin{eqnarray}\label{b9}
u^{\underline{a}}_{\underline{m}}(\G^{\underline{m}})_{\underline{\mu\nu}}=
v^{\underline{\a}}_{\underline{\mu}}(\G^{\underline{a}})_{\underline{\a\b}}
v^{\underline{\b}}_{\underline{\nu}}
\end{eqnarray}
Eq. \p{b9} is a generalization \cite{ghs,bzst,bzp} of Cartan -- Penrose
representation \cite{rindler,penrose} of light~--~like vector (see Eqs.
\p{cp}, \p{cp1}, \p{a5}, \p{a6} below).

Eq. \p{b9} can be regarded as a manifestation of the composed nature of
vector harmonic variables. From the other hand $u$ and $v$ are the
representations of the same rotations and, thus, we have a possibility to
consider our system in both terms of $u$ variables or $v$ ones whenever it
is useful.

In the framework of geometrical approach \cite{bpstv,geo}
$u_{\underline{a}}^{\underline{m}}$ should be considered as a local moving
frame attached to each point of worldvolume supersurface and
$v_{\underline{\a}}^{\underline{\mu}}$ should be regarded \cite{bzst,bzp}
as a generalization of Newman -- Penrose dyades \cite{newman}.
Then an infinitesimal transport shall change the moving frame
variables by Lorentz rotation parametrized by $so(1,D-1)$ valued Cartan
forms $\Om^{\underline{ab}}=-\Om^{\underline{ba}}$
\begin{eqnarray}\label{b14}
du^{\underline{b}}_{\underline{m}}={u_{\underline{m}}}^{\underline{a}}
{\Om_{\underline{a}}}^{\underline{b}};\ \ \ \ \
dv^{\underline{\a}}_{\underline{\mu}}={1\over
4}\Om^{\underline{a}\underline{b}}v^{\underline{\b}}_{\underline{\mu}}
{(\G_{\underline{ab}})_{\underline{\b}}}^{\underline{\a}};\ \ \ \ \
\Om^{\underline{ab}}=u^{\underline{a}}_{\underline{m}}d
u^{\underline{bm}}
\end{eqnarray}

With taking into account Eqs. \p{rot} and \p{b9},
$$
u^{\underline{a}}_{\underline{m}}=(u^{\pm\pm}_{\underline{m}},
u^{i}_{\underline{m}}),\ \ \ {v_{\underline{\mu}}}^{\underline{\a}}=
(v^{+}_{\underline{\mu} q},v^{-}_{\underline{\mu}\dot q}),\ \ \
{v_{\underline{\a}}}^{\underline{\mu}}=(v^{-\underline{\mu}}_{q},
v^{+\underline{\mu}}_{\dot q}),\ \ \ v^{\underline{\mu}}_{\underline{\a}}
v^{\underline{\b}}_{\underline{\mu}}=\d_{\underline{\a}}^{\underline{\b}}
$$
it is naturally to split Cartan form
$\Om^{\underline{ab}}$ \p{b14} into the set of:
\begin{itemize}
\item
$SO(1,1)$ connection
\begin{eqnarray}\label{b16}
\Om^{(0)}(d)\equiv{{1\over2}u^{--}_{\underline{m}}du^{++\underline{m}}=
-{2\over D-2}v^{+\underline{\mu}}_{\dot q}dv^{-}_{\underline{\mu}\dot
q}={2\over D-2}v^{-\underline{\mu}}_{q}dv^{+}_{\underline{\mu}q}}
\end{eqnarray}
\item
$SO(D-2)$ connection
\begin{eqnarray}\label{b17}
\Om^{ij}(d)\equiv{u^{i}_{\underline{m}}du^{j\underline{m}}=
{2\over D-2}v^{-\underline{\mu}}_{q}\g^{ij}_{qp}dv^{+}_{\underline{\mu}p}
={2\over D-2}v^{+\underline{\mu}}_{\dot
q}\tilde{\g}^{ij}_{\dot{qp}}dv^{-}_{\underline{\mu}\dot p} }
\end{eqnarray}
\item
$SO(1,D-1)/(SO(1,1)\otimes{SO(D-2)})$ vielbeins
\begin{equation}\label{b18}
\Om^{++i}\equiv{u^{++}_{\underline{m}}du^{i\underline{m}}={2\over D-2}
v^{+\underline{\mu}}_{\dot
q}\tilde{\g}^{i}_{\dot{q}q}dv^{+}_{\underline{\mu}q} }
\end{equation}
\begin{equation}\label{b19}
\Om^{--i}\equiv{u^{--}_{\underline{m}}du^{i\underline{m}}={2\over D-2}
v^{-\underline{\mu}}_{q}\g^{i}_{q\dot{q}}dv^{-}_{\underline{\mu}\dot q}}
\end{equation}
\end{itemize}

By definition $\Om^{\underline{ab}}$ satisfies the Maurer -- Cartan
equations
\begin{equation}\label{b21}
d\Om^{\underline{ab}}-{\Om^{\underline{a}}}_{\underline{c}}
\Om^{\underline{cb}}=0
\end{equation}
(which are the integrability conditions for
\p{b14}). Having in mind the $SO(1,1)\otimes SO(D-2)$ invariant
decomposition \p{b16} -- \p{b19} we can split \p{b21} into
the following set of equations:
\begin{equation}\label{b22}
{\cal D}\Om^{++i}\equiv{d\Om^{++i}-\Om^{++i}\Om^{(0)}+\Om^{++j}\Om^{ji}=0}
\end{equation}
\begin{equation}\label{b23}
{\cal D}\Om^{--i}\equiv{d\Om^{--i}+\Om^{--i}\Om^{(0)}+\Om^{--j}\Om^{ji}=0}
\end{equation}
\begin{equation}\label{b24}
{\cal F} \equiv d \Om^{(0)} = {1\over 2} \Om^{--i} \Om^{++i}
\end{equation}
\begin{equation}\label{b25}
R^{ij}  \equiv d \Om^{ij} + \Om^{ik} \Om^{kj} =
- \Om^{--[i} \Om^{++j]}
\end{equation}
In the case of bosonic p -- branes with $p\geq 1$ Eqs. \p{b22} -- \p{b25}
give rise to Peterson~--~Codazzi, Gauss and Ricci equations of surface
theory respectively (see \cite{geo} and Refs. therein).

\subsection{Symmetries and equations of motion.}

To investigate the gauge symmetries of the action \p{c1} and to obtain
the equations of motion, let us consider the variation of the
functional \p{c1} with respect to the superfield variables (modulo
boundary terms):
\begin{equation}\label{c3}
\d{S}=\int_{{\cal{M}}^1}\,\,\d\rho^{++}E^{--}+\int_{{\cal{M}}^1}\,\,
\rho^{++}(d\Pi^{\underline{m}}(\d)u^{--}_{\underline{m}}-2id\Th\g^
{\underline{m}}\d\Th{u^{--}_{\underline{m}}}+\Pi^{\underline{m}}\d
{u^{--}_{\underline{m}}})
\end{equation}
$$
=\int_{{\cal{M}}^1}\,\,(\d\rho^{++}-\rho^{++}\Om^{(0)}(\d))E^{--}+
\int_{{\cal{M}}^1}\,\,\rho^{++}\Om^{--i}(\d)E^{i}
$$
$$
-\int_{{\cal{M}}^1}\,\,\rho^{++}E^{i}(\d)\Om^{--i}(d)
-4i\int_{{\cal{M}}^1}\,\,\rho^{++}E^{-\dot q}E^{-\dot q}(\d)
$$
\begin{equation}\label{c4}
-\int_{{\cal{M}}^1}\,\,E^{--}(\d)D\rho^{++}
\end{equation}
To get manifestly supersymmetric expression, the supersymmetric invariant
variation
$$
\Pi^{\underline{m}}(\d)=\d{X}^{\underline{m}}-i\d\Th\g^{\underline{m}}\Th
$$
is used in \p{c3} instead of $\d{X}^{\underline{m}}$:
\begin{eqnarray}\label{c5}
E^{--}(\d)=\Pi^{\underline{m}}(\d)u^{--}_{\underline{m}};
\ \ \ \ \
E^{i}(\d)=\Pi^{\underline{m}}(\d)u^{i}_{\underline{m}};\ \ \ \ \
E^{-\dot q}(\d)\equiv{\d\Th^{\underline{\m}}v_{\underline{\m}}^{-\dot q}}
\end{eqnarray}
In accordance with \p{b14} variations of moving frame variables are
described by Cartan forms \p{b16} -- \p{b19} dependent on the variation
symbol $\d$ instead of $d$. Thus, $\Om^{(0)}(\d)$ and $\Om^{--i}(\d)$
are the parameters of $SO(1,1)$ and $S^{D-2}\simeq
SO(1,D-1)/(SO(1,1)\otimes{SO(D-2)}\otimes{K_{D-2}})$ \cite{ghs}
transformations respectively. The absence of the parameters
$\Om^{ij}(\d)$, $\Om^{++i}(\d)$ of $SO(D-2)$ and $K_{D-2}$ transformations
reflects the gauge symmetries of the theory (see below).

In \p{c3}, \p{c4} and below the covariant derivative
$$
D\rho^{++}=d\rho^{++}-\rho^{++}\Om^{(0)}(d)
$$
$$
DE^{--}(\d)\equiv{dE^{--}(\d)+E^{--}(\d)\Om^{(0)}(d)}
$$
$$
DE^{i}=dE^{i}+E^{j}\Om^{ij},\ \ \dots
$$
is defined by spin ($SO(1,1)$) connection $\Om^{(0)}(d)$ and $SO(D-2)$
gauge fields $\Om^{ij}(d)$ (i.e. by pullback of \p{b16}, \p{b17}) being
induced by embedding.

From \p{c4} it is easy to see that, in addition to N=1 target
space supersymmetry, the action \p{c1} possesses the following symmetries
\footnote{For the reader convenience, we present below the
transformations of both $u$ and $v$ variables, though they are completely
determined by the expressions for Cartan forms
$\Om^{\underline{ab}}(\d)$ in accordance with Eqs. \p{b14}, \p{b16} --
\p{b19}.}:
\begin{itemize}
\item[1)]
$SO(1,1)$ symmetry
\begin{equation}\label{c8}
\Om^{(0)}(\d)=2\a;\ \ \
\d\rho^{++}=2\a\rho^{++};\ \ \
\d u^{++}_{\underline{m}}=2\a u^{++}_{\underline{m}};\ \ \
\d u^{--}_{\underline{m}}=-2\a u^{--}_{\underline{m}}
\end{equation}
$$
\d v^{+}_{\underline{\mu} q}=\a v^{+}_{\underline{\mu} q};\ \ \ \ \
\d v^{-}_{\underline{\mu}\dot q}=-\a v^{-}_{\underline{\mu}\dot q}
$$
\item[2)]
The symmetry under $SO(D-2)$ rotations
$$
\Om^{ij}(\d)=\a^{ij};\ \ \
\d u^{i}_{\underline{m}}=-u^{j}_{\underline{m}}\a^{ji};\ \ \
\d v^{+}_{\underline{\mu}q}=-{1\over 4}\a^{ij}v^{+}_{\underline{\mu}p}
\g^{ij}_{pq};\ \ \
\d v^{-}_{\underline{\mu}\dot q}=-{1\over
4}\a^{ij}v^{-}_{\underline{\mu}\dot p}{\tilde\g}^{ij}_{\dot p\dot q}
$$
reflecting the absence of $u^{i}_{\underline{m}}$
variables in the action \p{c1} (but $u^{i}_{\underline{m}}$ are present
indirectly due to the orthogonality constraint \p{a} on
$u^{++}_{\underline{m}}$ variables
$u^{++}_{\underline{m}}u^{i\underline{m}}=0$). It manifests itself in
absence of $\Om^{ij}(\d)$ variation in \p{c4}.
\item[3)]
$K_{D-2}$ "boost" symmetry \cite{ghs,bhbz0} arising from the absence of
$v^{+}_{\underline{\m} q}$ (or $u^{++}_{\underline{m}}$,
$u^{i}_{\underline{m}}$) variables in the action \p{c1}
\begin{equation}\label{c91}
\Om^{++i}(\d)=k^{++i};\ \ \
\d u_{\underline{m}}^{++}=u^{i}_{\underline{m}}k^{++i};\ \ \
\d u_{\underline{m}}^{i}={1\over 2}u^{--}_{\underline{m}}k^{++i};\ \ \
\d v^{+}_{\underline{\mu} q}={1\over 2}k^{++i}\g^{i}_{q\dot q}
v^{-}_{\underline{\mu}\dot q}
\end{equation}
\item[4)]
$n=D-2$ local worldline fermionic symmetry realized nonlinearly as
superfield irreducible $\k$~--~symmetry \cite{k,pashnev}
\begin{equation}\label{c10}
\d\Th^{\underline{\mu}}v_{\underline{\mu}q}^{+}=\k^{+q};\ \ \ \ \
\d{X^{\underline{m}}}=i\d\Th\g^{\underline{m}}\Th
\end{equation}
\item[5)]
Bosonic superfield $b$ -- symmetry \cite{k,pashnev,bnsv} (related to
reparametrization symmetry)
\begin{equation}\label{c9}
\d{X^{\underline{m}}}=E^{++}(\d)u^{--{\underline{m}}}\equiv{
b^{++}(\tau,\eta^{+q})u^{--\underline{m}}},\ \ \ \ \
\d\Th^{\underline{\mu}}=0.
\end{equation}
\end{itemize}
which close the algebra of the $\k$ -- symmetry transformations.

All the symmetries of the proposed action \p{c1}
are realized in completely irreducible way
\footnote{Pure component counterparts of all the considered symmetries are
present in Lorentz -- harmonic formulation of superparticles and null --
super -- p -- branes \cite{bhbz0}.}
(in contrast to the previously known superfield formulations, especially
in the case of D=10 superparticle \cite{gs92}). Some of them will be
used below for establishing a connection to the usual superfield
versions \cite{stv} -- \cite{bnsv}.

Equations of motion
\footnote{We would like to point out the fact that equations of motion
derived from the action can be splitted into the two different parts:
proper (dynamical) equations of motion and "rheotropic relations"
\cite{bsv} establishing a connection between target -- space and worldline
vielbeins.}
derived by variation over $\d\rho^{++}$,
$\Om^{--i}(\d)$, $E^{--}(\d)$,
$E^{-\dot q}(\d)=\d\Th^{\underline{\m}}v^{-}_{\underline{\m}\dot q}$ and
$E^{i}(\d)$ have the following form:
\begin{equation}\label{c12}
E^{--}=0,\ \ \ \ \ \rho^{++}E^{i}=0
\end{equation}
\begin{equation}\label{c14}
\rho^{++}E^{-\dot
q}\equiv{\rho^{++}d\Th^{\underline{\m}}v_{\underline{\m}}^{-\dot q}}=0
\end{equation}
\begin{equation}\label{c141}
\Om^{--i}(d)=0
\end{equation}
\begin{equation}\label{c13}
D\rho^{++}\equiv{d\rho^{++}-\rho^{++}\Om^{(0)}(d)}=0
\end{equation}

The variation with respect to the surface ${\cal{M}}^1$ (${\d S\over
\d\eta^{q}(\tau)}=0$) does not result in independent equation of motion.
This reflects superdiffeomorphism invariance of the generalized
action \cite{bsv} (see also subsection 2.4) and provides the possibility
to treat Eqs. \p{c12}~--~\p{c13} as superfield equations, which hold on
the whole worldline superspace $\Sigma^{(1\vert D-2)}$ \cite{bsv}.

Supposing $\rho^{++}\ne 0$ we can solve \p{c12} as
\begin{equation}\label{c15}
\Pi^{\underline{m}}\equiv dX^{\underline{m}}-id\Th\G^{\underline{m}}\Th
={1\over2}e^{++}u^{--{\underline{m}}}
\end{equation}
and identify by this a pullback of the target -- space one form
$E^{++}=\Pi^{\underline{m}}u^{++}_{\underline{m}}$ with worldline
"einbein" $e^{++}$
\begin{equation}\label{c151}
E^{++}=e^{++}
\end{equation}
This choice is possible due to the absence of worldline supereinbein one
-- form in the original action.

It shall be stressed, that due to \p{b9}
\begin{equation}\label{cp}
u^{--}_{\underline{m}}\d_{\dot p\dot q}=v^{-}_{\underline{\mu}\dot q}
\G_{\underline{m}}v^{-}_{\underline{\mu}\dot p}\Longrightarrow
u^{--}_{\underline{m}}={1\over D-2}v^{-}_{\underline{\mu}\dot q}
\G_{\underline{m}}v^{-}_{\underline{\mu}\dot p}
\end{equation}
and, hence \p{c15} reproduces the twistor -- like solution
\begin{equation}\label{cp1}
\Pi^{\underline{m}}_{++}\equiv D_{++}X^{\underline{m}}
-iD_{++}\Th\G^{\underline{m}}\Th={1\over D-2}
v^{-}_{\underline{\mu}\dot q}\G_{\underline{m}}
v^{-}_{\underline{\mu}\dot p}
\end{equation}
of the mass shell condition
$$
\Pi^{\underline{m}}_{++}\Pi_{++\underline{m}}=0
$$

Moreover, the fact that
$\Pi^{\underline{m}}$ is expressed in terms of only vector
vielbein $e^{++}$, while $d=e^{++}D_{++}+e^{+q}D_{+q}$, means that we have
obtained the "geometrodynamical" condition \cite{gs92}, \cite{gs2}
\begin{equation}\label{emb}
\Pi^{\underline{m}}_{+q}\equiv
D_{+q}X^{\underline{m}}-iD_{+q}\Th\G^{\underline{m}}\Th=0
\end{equation}
being the starting point of the standard doubly supersymmetric approach
\cite{stv,stvz,gs92,gs2}, as equation of motion without any use of
Lagrange multipliers.

From equation \p{c14} for $\Th^{\underline{\mu}}$ variables
\begin{equation}\label{c18}
\rho^{++}d\Th^{\underline{\mu}}v^{-}_{\underline{\mu}\dot{q}}=0
\end{equation}
we get rheotropic relation \cite{bsv}
\begin{equation}\label{c19}
D_{+q}\Th^{\underline{\mu}}v^{-}_{\underline{\mu}\dot{q}}=0
\end{equation}
as well as the proper equation of motion
\begin{equation}\label{c20}
D_{++}\Th^{\underline{\mu}}v^{-}_{\underline{\mu}\dot{q}}=0
\end{equation}

Below we will prove that \p{c20} is the only independent
dynamical superfield equation of superparticle theory, because Eq.
\p{c141} results from \p{c20}.

Eq. \p{c19} means that
$$
D_{+q}\Th^{\underline{\mu}}={A_{q}}^{p}v^{-\underline{\mu}}_{p}
$$
with an arbitrary matrix ${A_{q}}^{p}$ supposed to be non -- degenerate
(i.e. $det\ A\ne 0$).

Since $e^{+q}$ variables are not involved into construction of our action,
we can choose it to be induced by embedding, i.e. we can suppose
$$
E^{+q}=d\Th^{\underline{\mu}}v^{+}_{\underline{\mu}q},
$$
as we have done it for $e^{++}$ in \p{c151}. To make this point clear, let
us note that, due to the absence of $e^{+q}$ in the action \p{c1},
we have a symmetry \cite{bsv}
\begin{equation}\label{c21}
e^{+q}\mapsto \tilde{e}^{+q}=(e^{+p}+e^{++}\chi_{++}^{+p}){W_{p}}^{q},
\ \ \ \ \ detW\not= 0
\end{equation}
$$
\Longrightarrow D_{+q}\mapsto \tilde{D}_{+p}{W^{-1}_{q}}^{p};\ \ \ \
D_{++}\mapsto \tilde{D}_{++}-\chi^{+p}_{++}\tilde{D}_{+p}
$$
which allows us to obtain the relation ${A_{q}}^{p}={\d_{q}}^{p}$, or
\begin{equation}\label{c22}
D_{+q}\Th^{\underline{\mu}}=v^{\underline{\mu}}_{+q}
\end{equation}
as well as
\begin{equation}\label{35prime}
D_{++}\Th^{\underline{\mu}}v_{\underline{\mu}q}^{+}=0.
\end{equation}

So Eq. \p{c22} identifies the generalized twistor superfield
$D_{+q}\Th^{\underline{\mu}}$ \cite{stv} with spinor harmonic
$v^{-\underline{\mu}}_{q}$ \cite{bhbz0,ghs}, \cite{bzst} -- \cite{bzp} (see
\cite{bpstv} for super -- p -- brane case).

Eqs. \p{c22}, \p{35prime} can be represented by the one -- form equations
\begin{equation}\label{c23}
E^{+q}\equiv{d\Th^{\underline{\mu}}v_{\underline{\mu}}^{+q}=e^{+q}},\ \ \
\ \
E^{-\dot q}\equiv{d\Th^{\underline{\mu}}v_{\underline{\mu}\dot
q}^{-}=e^{++}D_{++}\Th^{\underline{\mu}}v_{\underline{\mu}\dot q}^{-}\equiv
{e^{++}\psi_{++\dot q}^{-}}}
\end{equation}
Henceforth, a pullback of target -- space Grassmann one -- form looks like
\begin{equation}\label{c24}
d\Th^{\underline{\mu}}=e^{+q}v^{-{\underline{\mu}}}_{q}+e^{++}\psi_{++
\dot{q}}^{-}v^{+{\underline{\mu}}}_{\dot q}
\end{equation}
and equation of motion \p{c20} means
$$
\psi^{-}_{++\dot q}=0
$$.

Thus, we have obtained the set of rheotropic conditions \p{c12},
\p{c15}, \p{c24} and the proper equations of motion \p{c13}, \p{c141} and
\p{c20}.

All these relations are written in terms of differential forms, so it is
easy to investigate their integrability conditions.

\subsection{"Off -- shell" description of N=1 superparticle.}

Let us, for a time, omit from the consideration the proper dynamical
equations of motion \p{c20} (and \p{c141}) as well as Eq. \p{c13} for
Lagrange multiplier superfield $\rho^{++}$ and study the integrability
condition for rheotropic relations \p{c12}, \p{c22}, \p{c151}
written in the form \p{c15}, \p{c24}.  They are \footnote{Remind that we
use external differential and external product of forms, so $dd=0$ and
$\Om_{S}\Om_{R}=(-)^{SR}\Om_{R}\Om_{S}$,
$d(\Om_{R}\Om_{S})=\Om_{R}d\Om_{S}+(-)^{S}d\Om_{R}\Om_{S}$ for products of
any bosonic $R$ -- and $S$ -- forms $\Om_{R}=dZ^{\underline{M}_{R}}\dots
dZ^{\underline{M}_{1}}\Om_{\underline{M}_{1}\dots \underline{M}_{R}}$
and
$\Om_{S}=dZ^{\underline{M}_{S}}\dots
dZ^{\underline{M}_{1}}\Om_{\underline{M}_{1}\dots \underline{M}_{S}}$.}
$$
d\Pi^{\underline{m}}\equiv -id\Th\G^{\underline{m}}d\Th={1\over 2}
e^{++}Du^{--\underline{m}}+{1\over 2}De^{++}u^{--\underline{m}}
$$
\begin{equation}\label{a1}
={1\over 2}e^{++}\Om^{--i}u^{i\underline{m}}+{1\over 2}
T^{++}u^{--\underline{m}}
\end{equation}
$$
0 \equiv dd\Th^{\underline{\mu}}=De^{+q}v^{-\underline{\mu}}_{q}+
De^{++}\psi^{-}_{++\dot q}v^{+\underline{\mu}}_{\dot q}
+e^{++}D\psi^{-}_{++\dot q}v^{+\underline{\mu}}_{\dot q}+e^{+q}D
v^{-\underline{\mu}}_{q}+e^{++}\psi^{-}_{++\dot
q}Dv^{+\underline{\mu}}_{\dot q}
$$
\begin{equation}\label{a2}
=(T^{+q}-{1\over 2}e^{++}\land \Om^{++i}\g^{i}_{q\dot q}\psi^{-}_{++\dot
q})v^{-\underline{\mu}}_{q}
+(T^{++}\psi^{-}_{++\dot q}+e^{++}D\psi^{-}_{++\dot q}-{1\over 2}e^{+q}
\Om^{--i}\g^{i}_{q\dot q})v^{+\underline{\mu}}_{\dot q}
\end{equation}
where the covariant differential $D$ involves $SO(1,1)$ and $SO(D-2)$
connections induced by embedding, i.e. coincident with the pullbacks of
the forms $\Om^{(0)}(d)$ and $\Om^{ij}(d)$. The later fact can be
represented by the following relations
$$
\Om^{(0)}(D)=0;\ \ \ \ \ \Om^{ij}(D)=0
$$

As a result, covariant differentials of harmonic variables involve the
$SO(1,D-1)/SO(1,1)\otimes SO(D-2)$ forms only (compare with \p{b14}). To
make this point clear let us write the explicit expressions for D=10 case,
where $\g^{i}_{q\dot q}$ is the $SO(8)$ $\g$ -- matrices,
$\tilde{\g}^{i}_{q\dot q}=\g^{i}_{q\dot q}$,
$\g^{ij}=\g^{[i}\tilde{\g}^{j]}$,
$\tilde{\g}^{ij}=\tilde{\g}^{[i}\g^{j]}$:
$$
Dv^{-\underline{\mu}}_{q}\equiv dv^{-\underline{\mu}}_{q}+{1\over 2}
\Om^{(0)}v^{-\underline{\mu}}_{q}-{1\over 4}\Om^{ij}\g^{ij}_{qp}
v^{-\underline{\mu}}_{p}=-{1\over 2}\Om^{--i}\g^{i}_{q\dot q}
v^{+\underline{\mu}}_{\dot q}
$$
$$
Dv^{+\underline{\mu}}_{\dot q}\equiv dv^{+\underline{\mu}}_{\dot
q}-{1\over 2}\Om^{(0)}v^{+\underline{\mu}}_{\dot q}-{1\over
4}\Om^{ij}\tilde{\g}^{ij}_{\dot q\dot p}
v^{+\underline{\mu}}_{\dot p}=-{1\over 2}\Om^{++i}\g^{i}_{q\dot q}
v^{-\underline{\mu}}_{q}
$$
\begin{equation}\label{a22}
Du^{--}_{\underline{m}}\equiv du^{--}_{\underline{m}}+\Om^{(0)}
u^{--}_{\underline{m}}=u^{i}_{\underline{m}}\Om^{--i},\ \ \dots
\end{equation}

In Eqs. \p{a1}, \p{a2} $T^{++}$ and $T^{+q}$ are the bosonic and
fermionic torsion two -- forms
\begin{equation}\label{a3}
T^{++}\equiv De^{++}=de^{++}-e^{++}\Om^{(0)}
\end{equation}
\begin{equation}\label{a4}
T^{+q}\equiv De^{+q}=de^{+q}-{1\over 2}e^{+q}\Om^{(0)}+{1\over 4}
e^{+p}\Om^{ij}\g^{ij}_{pq}
\end{equation}

Contracting \p{a1} and \p{a2} with moving frame vectors ${1\over
2}u^{++}_{\underline{m}}, u^{i}_{\underline{m}}$ and spinor harmonics
$v^{+}_{\underline{\mu}q},v^{-}_{\underline{\mu}\dot q}$ we get:
\begin{equation}\label{a7}
T^{++}=-2iE^{+q}E^{+q}
\end{equation}
\begin{equation}\label{a8}
e^{++}\Om^{--i}=-4iE^{+q}\g^{i}_{q\dot q}E^{-\dot q}
\end{equation}
\begin{equation}\label{a9}
T^{+q}={1\over 2}e^{++}\land \Om^{++i}\g^{i}_{q\dot q}\psi^{-}_{++\dot
q}
\end{equation}
\begin{equation}\label{a10}
e^{+q}\Om^{--i}\g^{i}_{q\dot q}=2e^{++}D\psi^{-}_{++\dot q}+2T^{++}
\psi^{-}_{++\dot q}
\end{equation}
where the consequences of Eq. \p{b9}
\begin{equation}\label{a5}
u^{++}_{\underline{m}}\G^{\underline{m}}_{\underline{\mu\nu}}=
2v^{+}_{\underline{\mu}q}v^{+}_{\underline{\nu}q}
\end{equation}
\begin{equation}\label{a6}
u^{i}_{\underline{m}}\G^{\underline{m}}_{\underline{\mu\nu}}=
2v^{+}_{\{\underline{\mu}\vert q}\g^{i}_{q\dot
q}v^{-}_{\vert \underline{\nu}\}\dot q}
\end{equation}
have been used.

Taking into account the rheotropic conditions \p{c23} we can rewrite Eqs.
\p{a7} -- \p{a10} in the form
\begin{equation}\label{a11}
T^{++}=-2ie^{+q}e^{+q}
\end{equation}
\begin{equation}\label{a12}
e^{++}\Om^{--i}=-4ie^{++}e^{+q}\g^{i}_{q\dot q}\psi^{-}_{++\dot q}
\end{equation}
\begin{equation}\label{a111}
T^{+q}={1\over 2}e^{++}\land \Om^{++i}\g^{i}_{q\dot q}\psi^{-}_{++\dot q}
\end{equation}
\begin{equation}\label{a121}
e^{+q}\Om^{--i}\g^{i}_{q\dot q}=2e^{++}D\psi^{-}_{++\dot q}-4ie^{+q}e^{+q}
\psi^{-}_{++\dot q}
\end{equation}

Eq. \p{a11} contains the set of the worldline superspace torsion
constraints, in particular, the most essential one
\begin{equation}\label{a13}
{T_{+q+p}}^{++}=-4i\d_{qp}
\end{equation}

The appearance of the worldvolume superspace torsion
constraint as equations of motion of the generalized action
for super -- p -- branes was noted in \cite{bsv}.

Eq. \p{a111} gives the expression for Grassmann torsion two -- form. It
means that its component with Grassmann indices vanishes
\begin{equation}\label{a14}
{T_{+p+r}}^{+q}=0
\end{equation}
and component with one bosonic index is expressed in terms of
$\psi^{-}_{++\dot q}$ and spinor component of $\Om^{++i}$
\begin{equation}\label{a15}
{T_{++\ +p}}^{+q}=-{1\over 2}\Om^{++i}_{+p}\g^{i}_{q\dot
q}\psi^{-}_{++\dot q}
\end{equation}
Note that \p{a15}
and, hence, the whole Grassmann torsion two form vanishes on --
shell (see Eq. \p{c20} above).

Eq. \p{a121} with taking into account \p{a11} can be decomposed into the
following set of relations
\begin{equation}\label{a16}
\Om^{--i}_{\{+p}\g^{i}_{q\}\dot q}=-4i\d_{qp}\psi^{-}_{++\dot
q}
\end{equation}
\begin{equation}\label{a17}
2D_{+q}\psi^{-}_{++\dot q}=-\Om^{--i}_{++}\g^{i}_{q\dot q}
\end{equation}
The solution of Eq. \p{a16} has the form
\footnote{In D=10 the triality relations:
$$
\g^{i}_{\{q\vert \dot q}\tilde{\g}^{j}_{\dot q\vert p\}}=\d^{ij}
\d_{qp};\ \ \ \
\tilde{\g}^{i}_{\{\dot q\vert q}\g^{j}_{q\vert \dot p\}}=\d^{ij}
\d_{\dot q\dot p};\ \ \ \
\g^{i}_{\{q\vert \dot q}\g^{i}_{\vert p\}\dot p}=\d_{qp}
\d_{\dot q\dot p}
$$
shall be used.}
\begin{equation}\label{a18}
\Om^{--i}_{+q}=-4i\g^{i}_{q\dot q}\psi^{-}_{++\dot q}
\end{equation}
Hence, Eq. \p{a121} determines completely the form $\Om^{--i}$
\begin{equation}\label{a19}
\Om^{--i}=-4ie^{+q}\g^{i}_{q\dot q}\psi^{-}_{++\dot q}-{2\over D-2}
e^{++}D_{+q}\psi^{-}_{++\dot q}\g^{i}_{q\dot q}
\end{equation}
and puts the following restriction on the Grassmann superfield
$\psi^{-}_{++\dot q}$
\begin{equation}\label{a20}
D_{+q}\psi^{-}_{++\dot q}={1\over D-2}\g^{i}_{q\dot q}\g^{i}_{p\dot p}
D_{+p}\psi^{-}_{++\dot p}
\end{equation}

It should be note that Eq. \p{a12} as well as
the second Peterson -- Codazzi equation \p{b23} $D\Om^{--i}=0$ are
satisfied identically for the form \p{a19}, \p{a20}. $SO(1,1)$ and
$SO(D-2)$ curvatures of worldline superspace are defined by Gauss \p{b24}
and Ricci \p{b25} equations and involve the form $\Om^{++i}$, unrestricted
by rheotropic conditions, as the expressions \p{a9}, \p{a15} for
the Grassmann torsion two -- form do.

The restrictions on the form $\Om^{++i}$ follow from the first Peterson --
Codazzi equation only
$$
D\Om^{++i}=0\Longrightarrow e^{+q}D\Om^{++i}_{+q}+T^{++}\Om^{++i}_{++}
$$
\begin{equation}\label{a21}
+T^{+q}\Om^{++i}_{+q}+e^{++}D\Om^{++i}_{++}=0
\end{equation}
The independent consequence of Eq. \p{a21} is
\begin{equation}\label{a222}
D_{+\{p}\Om^{++i}_{+q\}}=2i\d_{pq}\Om^{++i}_{++}
\end{equation}
where Eqs. \p{a11}, \p{a111} are taken into account.
Thus,
$$
\Om^{++i}=(e^{+q}-{i\over 2(D-2)}e^{++}D_{+q})\Om^{++i}_{+q}
$$
with an arbitrary Grassmann superfield $\Om^{++i}_{+q}$.

Under the boost transformations \p{c91} this superfield transforms as
$$
\d\Om^{++i}_{+q}=D_{+q}k^{++i}
$$
However, this symmetry can not be used to gauge $\Om^{++i}_{+q}$ to zero,
except for the simplest case D=3.

Hence, on mass shell N=1 superparticle can be described by:
\begin{itemize}
\item[1)]
The covariant Cartan forms
\begin{equation}\label{aa1}
\Om^{++i}=(e^{+q}-{i\over 2(D-2)}e^{++}D_{+q})\Om^{++i}_{+q}
\end{equation}
\begin{equation}\label{aa2}
\Om^{--i}=-4ie^{+q}\g^{i}_{q\dot q}\psi^{-}_{++\dot q}-{2\over D-2}
e^{++}D_{+q}\psi^{-}_{++\dot q}\g^{i}_{q\dot q}
\end{equation}
where superfield $\psi^{-}_{++\dot q}$ is restricted by
\begin{equation}\label{aa3}
D_{+q}\psi^{-}_{++\dot q}={1\over D-2}\g^{i}_{q\dot q}\g^{i}_{p\dot p}
D_{+p}\psi^{-}_{++\dot p}.
\end{equation}
Superfield $\Om^{++i}_{+q}$ \p{aa1} satisfying
\begin{equation}\label{aa33}
D_{+\{q}\Om^{++i}_{+p\}}=\d_{pq}{1\over D-2}D_{+r}\Om^{++i}_{+r}
\end{equation}
transforms as follows
\begin{equation}\label{aa4}
\d\Om^{++i}_{+q}=D_{+q}k^{++i}
\end{equation}
under the action of (gauge) boost symmetry \p{c91}.
\item[2)]
Worldline supergravity $(e^{++},e^{+q})$ satisfying the constraints
\begin{equation}\label{aa5}
T^{++}=-2ie^{+q}e^{+q}
\end{equation}
\begin{equation}\label{aa6}
T^{+q}={1\over2}e^{++}e^{+p}\Om^{++i}_{+p}\g^{i}_{q\dot q}\psi^{-}_{++\dot
q}
\end{equation}
\item[3)]
Induced $SO(1,1)$, $SO(D-2)$ connections $\Om^{(0)}$, $\Om^{ij}$ whose
curvatures are determined by Gauss and Ricci equations
\begin{equation}\label{aa7}
{\cal F} \equiv d \Om^{(0)} = {1\over 2} \Om^{--i} \Om^{++i}
\end{equation}
\begin{equation}\label{aa8}
R^{ij}  \equiv d \Om^{ij} + \Om^{ik} \Om^{kj} =
- \Om^{--[i} \Om^{++j]}
\end{equation}
\end{itemize}

In the present consideration we have not take into account scalar
density superfield $\rho^{++}$. It can be considered as compensator of
$SO(1,1)$ gauge transformations.

\subsection{The action independence on the surface.}

In this subsection we explain the fact that the generalized action
possesses superdiffeomorphism invariance and variation with respect to the
surface ${\cal M}^1$ does not result in independent equations of motion.

First of all, let us note that the general variation of Lagrangian one --
form (including the surface ${\cal M}^1$ variation) is
\begin{equation}\label{ai1}
\d{\cal L}=(d{\cal L})(d,\d)+d({\cal L}(\d))
\end{equation}
or, in details,
$$
\d{\cal L}=-d\rho^{++}E^{--}(\d)+\d\rho^{++}E^{--}(d)+\rho^{++}E^{--}(\d)
\Om^{(0)}(d)
$$
$$
-\rho^{++}E^{--}(d)\Om^{(0)}(\d)-\rho^{++}E^{i}(\d)\Om^{--i}(d)+\rho^{++}
E^{i}(d)\Om^{--i}(\d)
$$
\begin{equation}\label{ai2}
-4i\rho^{++}E^{-\dot q}(d)E^{-\dot q}(\d)+d(\rho^{++}E^{--}(\d))
\end{equation}

The last term has not contribution into variation due to the chosen
initial conditions. On the surface of the rheotropic relations
\begin{equation}\label{ai3}
E^{--}=0;\ \ \ \ \ E^{i}=0
\end{equation}
the Lagrangian form transforms as follows:
$$
\d{\cal L}\vert_{rh.}=-d\rho^{++}E^{--}(\d)+\rho^{++}E^{--}(\d)
\Om^{(0)}(d)
$$
\begin{equation}\label{ai4}
-\rho^{++}E^{i}(\d)\Om^{--i}(d)
-4i\rho^{++}E^{-\dot q}(d)E^{-\dot q}(\d)
\end{equation}

To specify \p{ai4}, consider the variation of the action \p{c1} under the
general coordinate transformations and the change of the surface ${\cal
M}^1$
\begin{equation}\label{ai5}
\d_{g.c.+surf.}S=\int_{\d{\cal M}^1}\,\,{\cal L}=
\int_{{\cal M}^1}\,\,({\cal
L}(\tau+\d\tau,\eta^{q}(\tau)+\d\eta^{q}(\tau))- {\cal
L}(\tau,\eta^{q}(\tau)))=\int_{{\cal M}^1}\,\,\d_{\tau,\eta}{\cal L}
\end{equation}
where $\d_{\tau,\eta}{\cal L}$ is determined by Eq. \p{ai4} with
\begin{equation}\label{ai6}
E^{--}(\d)=\d\tau\Pi^{\underline{m}}_{\tau}u^{--}_{\underline{m}}+
\d\eta^{q}\Pi^{\underline{m}}_{q}u^{--}_{\underline{m}}
\end{equation}
\begin{equation}\label{ai7}
E^{i}(\d)=\d\tau\Pi^{\underline{m}}_{\tau}u^{i}_{\underline{m}}+
\d\eta^{q}\Pi^{\underline{m}}_{q}u^{i}_{\underline{m}}
\end{equation}
\begin{equation}\label{ai9}
E^{-\dot q}(\d)=\d\tau E^{-\dot q}_{\tau}+\d\eta^{q} E^{-\dot q}_{q}
\end{equation}
The first two variations \p{ai6}, \p{ai7} vanish due to the consequences
from the rheotropic relations \p{ai3}
\begin{equation}\label{ai8}
\Pi^{\underline{m}}_{\tau}=e^{++}_{\tau}u^{--}_{\underline{m}};\ \ \ \ \
\Pi^{\underline{m}}_{q}=e^{++}_{q}u^{--}_{\underline{m}}
\Longrightarrow \Pi^{\underline{m}}_{\tau}u^{--}_{\underline{m}}=0;
\ \ \ \ \ \Pi^{\underline{m}}_{q}u^{--}_{\underline{m}}=0
\end{equation}
and properties of vector harmonics \p{a}.

Further, the variation \p{ai9}
can not provide a new equation of motion, because it is involved
\p{ai4} in the combination with $E^{-\dot q}(d)$
\begin{equation}\label{ai10}
E^{-\dot q}(d)E^{-\dot q}(\d)=e^{+p}E^{-\dot q}_{+p}(2\d\tau E^{-\dot
q}_{\tau}+\d\eta^{q}E^{-\dot q}_{q})
\end{equation}
only. The latter vanishes due to the spinorian rheotropic relation
\begin{equation}\label{ai11}
E^{-\dot q}_{+p}\equiv
D_{+p}\Th^{\underline{\mu}}v^{-}_{\underline{\mu}\dot q}=0
\end{equation}

Hence
\begin{equation}\label{ind3}
\d_{g.c.+surf.}S=0
\end{equation}
holds as a result of the rheotropic conditions only.

All mentioned above make clear the condition of the
off -- shell superdiffeomorphism invariance in the rheonomic sense,
proposed in \cite{rheo} and considered in \cite{bsv}. This condition
require the vanishing of the external derivative of the Lagrangian form
\begin{equation}\label{***}
d{\cal L}=(d\rho^{++}-\rho^{++}\Om^{(0)})E^{--}+\rho^{++}E^{i}\Om^{--i}(d)
-i\rho^{++}d\Th\G^{\underline{m}}d\Th u^{--}_{\underline{m}}=0
\end{equation}
due to the rheotropic conditions only and the non -- dynamical meaning
the rheotropic conditions (i.e., the rheotropic conditions shall not
result in the proper dynamical equations). In this form the proof of the
superdiffeomorphism invariance, being equivalent to the presented above,
becomes much more short. Namely,
the first two terms in Eq. \p{***} vanish due to the rheotropic relations
\p{ai3}.  The last term (after the complete decomposition onto the
supervielbein forms) can be rewritten as follows:
$$
-i\rho^{++}e^{+q}(2e^{++}D_{++}\Th v^{-}_{\dot q}+e^{+q}D_{+q}\Th
v^{-}_{\dot q})D_{+q}\Th v^{-}_{\dot q}
$$
and vanishes due to rheotropic relation \p{ai11}.

This means the off -- shell superdiffeomorphism invariance of N=1
superparticle action in rheonomic sense \cite{rheo,bsv}, because, as it
have been proved in subsection 2.3, the rheotropic conditions do not lead
to equations of motion.

\subsection{"On -- shell" description of N=1 superparticle in diverse
dimensions.}

When the proper (dynamical) equation of motion \p{c20} is taken into
account
\begin{equation}\label{os1}
\psi^{-}_{++\dot
q}\equiv D_{++}\Th^{\underline{\mu}}v^{-}_{\underline{\mu}\dot{q}}=0
\end{equation}
we get from Eq. \p{a19}
\begin{equation}\label{os2}
\Om^{--i}=0
\end{equation}

Thus, we have proved the dependence of the equation of motion \p{c20}.

The curvatures of worldline superspace (see \p{aa7}, \p{aa8}) vanish due
to Eq. \p{os2}
\begin{equation}\label{os3a}
{\cal F} \equiv d \Om^{(0)}=0
\end{equation}
\begin{equation}\label{os3b}
R^{ij}  \equiv d \Om^{ij} + \Om^{ik} \Om^{kj} =0
\end{equation}

As a result, the induced connections $\Om^{(0)}$, $\Om^{ij}$ can be gauged
away and, thus, have not physical degrees of freedom.

Eqs. \p{os3a}, \p{os3b} mean that Peterson -- Codazzi equation \p{b22} can
be solved by
\begin{equation}\label{os4}
\Om^{++i}=Dl^{++i}
\end{equation}
Henceforth, superfield $l^{++i}$ transforms additively under the boost
symmetry \p{c91}
$$
\d l^{++i}=k^{++i}
$$
and can be gauged away
$$l^{++i}=0$$
In this gauge
\begin{equation}\label{os41}
\Om^{++i}=0
\end{equation}
and worldline supergravity $(e^{++},e^{+q})$ is characterized by
the flat constraint
\begin{equation}\label{os8}
T^{++}=-2ie^{+q}e^{+q};\ \ \ \ \ T^{+q}=0
\end{equation}

To get the general solution of \p{os8} let us consider the exact form of
$SO(1,1)$ and $SO(D-2)$ connections being the solution of Eqs. \p{c13},
\p{os3a} and \p{os3b} respectively:
\begin{equation}\label{80prime}
\Om^{(0)}={d\rho^{++}\over \rho^{++}}
\end{equation}
%Eq. \p{os3b} means that $\Om^{ij}$ can be presented in the form
\begin{equation}\label{80prime1}
\Om^{ij}=dG^{ik}G^{jk};\ \ \ \ \ G^{ik}G^{jk}=\d^{ij}
\end{equation}
In \p{80prime1} $G^{ik}$ are arbitrary orthogonal matrices.

Further it is convenient to use the spinor representation $\Om^{qp}$ for
$SO(D-2)$ connection, which is defined by the relations
\begin{equation}\label{80prime2}
\Om^{ij}(d)={2\over D-2}v^{-\underline{\mu}}_{q}\g^{ij}_{qp}
dv^{+}_{\underline{\mu}p}\equiv {2\over D-2}\g^{ij}_{qp}\Om^{qp}
\end{equation}
$$
\Om^{qp}={i\over 4}(\g_{ij})_{qp}\Om^{ij};\ \ \ \ \
R^{qp}={i\over 4}(\g_{ij})_{qp}R^{ij}=0;
$$
\begin{equation}\label{80prime3}
\Om^{qp}=dG^{qr}G^{pr};\ \ \ \ \ G^{qr}G^{pr}=\d^{qp}
\end{equation}

Using Eqs. \p{80prime}, \p{80prime3}, we can write down the second
Eq. \p{os8}:
$$
T^{+q}\equiv De^{+q}=de^{+q}-{1\over2}e^{+q}\Om^{(0)}+e^{+p}\Om^{pq}
$$
$$
=de^{+q}+\sqrt{\rho^{++}}e^{+q}d({1\over \sqrt{\rho^{++}}})+e^{+p}
dG^{pk}G^{qk}
$$
\begin{equation}\label{80prime4}
=\sqrt{\rho^{++}}G^{qk}d({e^{+p}G^{pk}\over \sqrt{\rho^{++}}})=0
\end{equation}
The later equation can be solved by
\begin{equation}\label{80prime5}
e^{+p}=\sqrt{\rho^{++}}G^{pq}d{\tilde{\eta}}^{q}
\end{equation}
with arbitrary Grassmann function
${\tilde{\eta}}^{q}={\tilde{\eta}}^{q}(\tau,\eta^{q})$.

Henceforth, the first equation from \p{os8} becomes
\begin{equation}\label{80prime6}
T^{++}\equiv De^{++}=de^{++}-e^{++}\Om^{(0)}=\rho^{++}d({e^{++}
\over \rho^{++}})=-2i\rho^{++}d{\tilde{\eta}}^{q}d{\tilde{\eta}}^{q}
\end{equation}
and has the natural solution
\begin{equation}\label{80prime7}
e^{++}=\rho^{++}(d\tilde{\tau}-2id{\tilde{\eta}}^{q}{\tilde{\eta}}^{q});
\ \ \ \ \ \tilde{\tau}=\tilde{\tau}(\tau,\eta^{q})
\end{equation}
With taking into account \p{80prime5}, \p{80prime7} and \p{os1}, Eqs.
\p{c15}, \p{c24} acquire the form
\begin{equation}\label{80prime8}
dX^{\underline{m}}-id\Th\G^{\underline{m}}\Th={1\over 2}\rho^{++}
(d\tilde{\tau}-2id{\tilde\eta}^{q}{\tilde\eta}^{q})u^{--\underline{m}}
\end{equation}
\begin{equation}\label{80prime9}
d\Th^{\underline{\mu}}=\sqrt{\rho^{++}}G^{qp}d{\tilde\eta}^{p}
v^{-\underline{\mu}}_{q}
\end{equation}
Using the evident relation
$$
d{\tilde\eta}^{q}=d\tau\partial_{\tau}{\tilde\eta}^{q}+d{\eta}^{+p}
D_{+p}{\tilde\eta}^{q}
$$
we get from \p{80prime9}
\begin{equation}\label{80prime10}
D_{+p}\Th^{\underline{\mu}}=\sqrt{\rho^{++}}G^{rq}D_{+p}{\tilde\eta}^{q}
v^{-\underline{\mu}}_{r}
\end{equation}
\begin{equation}\label{80prime11}
\partial_{\tau}\Th^{\underline{\mu}}=\sqrt{\rho^{++}}G^{qp}\partial_{\tau}
{\tilde\eta}^{p}v^{-\underline{\mu}}_{q}
\end{equation}

Eq. \p{80prime11} leads to the standard equation of motion for Brink --
Schwarz superparticle
\begin{equation}\label{80prime13}
\Pi^{\underline{m}}_{\tau}(\G_{\underline{m}})_{\underline{\mu\nu}}
\partial_{\tau}\Th^{\underline{\mu}}=0
\end{equation}
because, due to \p{80prime8},
\begin{equation}\label{80prime12}
\Pi^{\underline{m}}_{\tau}\equiv
\partial_{\tau}X^{\underline{m}}-i\partial_{\tau}\Th\G^{\underline{m}}
\Th={1\over2}\rho^{++}u^{--\underline{m}}(\partial_{\tau}\tilde\tau-
2i\partial_{\tau}{\tilde\eta}^{q}{\tilde\eta}^{q})
\end{equation}
and
$$
u_{\underline{m}}^{--}\G^{\underline{m}}_{\underline{\mu\nu}}
v_{q}^{-\underline{\nu}}
=2v_{\underline{\mu}\dot q}^{-}v_{\underline{\nu}\dot q}^{-}
v_{q}^{-\underline{\nu}}=0
$$

Using the general coordinate transformations in worldline superspace
we can fix the gauge $\tilde{\tau}=\tau$, $\tilde{\eta}^{q}=\eta^{q}$,
which is conserved by superconformal transformations. In this gauge
$$
e^{+q}=\sqrt{\rho^{++}}d\eta^{q};\ \ \ \ \ \
e^{++}=\rho^{++}(d\tau-2id\eta^{q}\eta^{q})
$$
From the other hand, using Eqs. \p{os3a} and \p{80prime}, we can fix the
gauge $\Om^{(0)}=0$ by imposing $\rho^{++}=const$, and all the Cartan
forms vanish
$$
\Om^{\underline{ab}}=0
$$

In this gauge harmonics $u$ and $v$ are the constant vectors and spinors
constrained by \p{a}, \p{aa} respectively
$$
du=0;\ \ \ \ \ dv=0
$$
R.h.s. of Eqs. \p{c15}, \p{c24} becomes very simple
$$
dX^{\underline{m}}-id\Th\G^{\underline{m}}\Th={1\over 2}
(d\tau-2id\eta^{q}\eta^{q})u^{--\underline{m}}
$$
$$
d\Th^{\underline{\mu}}=d\eta^{q}v^{-\underline{\mu}}_{q}
$$
and can be used to extract the solution of mass shell constraints
$$
{\dot x}^{\underline{m}}={1\over 2}u^{--\underline{m}}
={1\over 2(D-2)}
v^{-}_{q}\G^{\underline{m}}v^{-}_{q}\Longrightarrow
({\dot x}^{\underline{m}})^{2}=0
$$
as well as equations of motion (in the form, possessing only the global
woldline supersymmetry \cite{vz})
$$
\ddot{x}=0;\ \ \ \ \
{\dot\th}^{\underline{\mu}}=0
$$
for leading components of the superfields $X$ and $\Th$,
$$
x^{\underline{m}}=X^{\underline{m}}(\tau,\eta=0);\ \ \ \ \
\th^{\underline{\mu}}=\Th^{\underline{\mu}}\vert_{\eta=0}
$$
and expressions for their higher components in terms of $x$, $\th$ and
constant twistor -- like variables $v^{-}_{q}$
$$
D_{+q}X^{\underline{m}}\vert_{\eta=0}
=-2i\eta^{q}u^{--\underline{m}}+iv^{-}_{q}
\G^{\underline{m}}\th
$$
$$
D_{+q}\Th^{\underline{\mu}}\vert_{\eta=0}=v^{-\underline{\mu}}_{q}
$$
etc.

Hence, we have proved the classical equivalence of N=1 superparticle
description in the framework of the generalized action principle approach
with standard one.

\subsection{Action in analytical basis.}

In the analytical coordinate basis ${\cal Z}^{\underline{\cal A}}=
({\tilde Z}^{\underline{A}},v^{\underline{\a}}_{\underline{\mu}})=
(X^{++},X^{--},{\tilde X}^{i},\Th^{\underline{\a}},
v^{\underline{\a}}_{\underline{\mu}})$
$$
X^{\underline{a}}\equiv
X^{\underline{m}}u^{\underline{a}}_{\underline{m}}=(X^{++},X^{--},X^{i})=
(X^{\underline{m}}u^{++}_{\underline{m}},X^{\underline{m}}
u^{--}_{\underline{m}},X^{\underline{m}}u^{i}_{\underline{m}})
$$
$$
\Th^{\underline{\a}}\equiv \Th^{\underline{\mu}}
v^{\underline{\a}}_{\underline{\mu}}=(\Th^{+q},\Th^{-\dot q})=
(\Th^{\underline{\mu}}v^{+q}_{\underline{\mu}},\Th^{\underline{\mu}}
v^{-\dot q}_{\underline{\mu}})
$$
$$
{\tilde X}^{i}=X^{i}+{i\over 4}\Th^{+}_{q}\g^{i}_{q\dot q}\Th^{-}_{\dot q}
$$
of a target superspace the action \p{c1} acquires the form
\begin{equation}\label{anal1}
S=\int_{{\cal{M}}^1}\,\,\rho^{++}(DX^{--}-2iD\Th^{-}_{\dot q}
\Th^{-}_{\dot q}-{\tilde X}^{i}\Om^{--i})
\end{equation}
where
\footnote{It shall be noted that the shift of the transverse components
similar to the \p{anal2} one plays an important role in the investigation
performed in \cite{sezgin}, where the linearized version of the formalism
\cite{bpstv} was developed and applied for a wide class of supersymmetric
extended objects, including D=11 super -- five brane.}
$$
DX^{--}\equiv dX^{--}+X^{--}\Om^{(0)}(d)
$$
$$
D\Th^{-\dot q}\equiv d\Th^{-\dot q}+{1\over 2}\Th^{-\dot q}\Om^{(0)}-
{1\over 4}\Om^{ij}{\tilde\g}^{ij}_{\dot q\dot p}\Th^{-\dot p}
$$
\begin{equation}\label{anal2}
{\tilde X}^{i}=X^{i}+{i\over 4}\Th^{+}_{q}\g^{i}_{q\dot q}
\Th^{-}_{\dot q}
\end{equation}

The point is important that the functional \p{anal1} does not involve the
variables $\Th^{+q}$ and $X^{++}$, which transform additively under
(irreducible superfield) $\k$ -- symmetry \p{c10} and $b$ -- symmetry
\p{c9} respectively (i.e. are the compensators of these symmetries).

Such type effect was found by E.Sokatchev
\cite{sok87} (in superparticle action written in Hamiltonian form with
use of vector Lorentz harmonic variables).

This observation seems to be important
for further attempts to find a way for superparticle
covariant quantization in the framework of the generalized action
approach.

\section{Coupling to Maxwell background field.}

The generalized action describing superparticle interacting
with Abelian background has the form
\begin{equation}\label{bac1}
S=\int_{{\cal{M}}^1}\,\,(\rho^{++}E^{--}+{\cal A})
\end{equation}
where, as in \p{c1}, ${\cal{M}}^1=\{(\tau,\eta^{q}):
\eta^{q}=\eta^{q}(\tau)\}$ is an arbitrary bosonic line in worldline
superspace. The superfield $\rho^{++}$ and the form $E^{--}$ are defined
in section 2.1 and
\begin{equation}\label{baca}
{\cal A}=dZ^{\underline{M}}{\cal A}_{\underline{M}}(X^{\underline{m}}(\tau,
\eta^{+q}),\Th^{\underline{\mu}}(\tau,\eta^{+q}));\ \ \ \ \
Z^{\underline{M}}=\{X^{\underline{m}},\Th^{\underline{\mu}}\}
\end{equation}

To obtain equations of motion we have to vary the action \p{bac1} with
taking into account the fact that a differential form
variation is similar to the Lie derivative operation, i.e. (see subsection
3.1 below)
\begin{equation}\label{bac3}
\d{\cal A}=(d{\cal A})(d,\d)+d({\cal A}(\d))
\end{equation}

Under assumption about trivial initial conditions, the contribution from
the last term of \p{bac3} vanishes and
\begin{equation}\label{bac4}
\int_{{\cal{M}}^1}\,\,\d{\cal A}{\equiv \int_{{\cal{M}}^1}\,\,F(d,\d)}=
\int_{{\cal{M}}^1}\,\,E^{\underline{B}}(d)E^{\underline{A}}(\d)
F_{\underline{AB}}
\end{equation}
where $F_{\underline{AB}}$ is the strength tensor superfield defined by
\begin{equation}\label{bac5}
d{\cal A}={1\over2}E^{\underline{B}}E^{\underline{A}}
F_{\underline{AB}}{\equiv F};\ \ \ \ \
F_{\underline{AB}}=-(-)^{\underline{AB}}F_{\underline{BA}}
\end{equation}

In Eqs. \p{bac4}, \p{bac5} $E^{\underline{A}}$ can be regarded as a set of
basic forms
$\Pi^{\underline{M}}=(\Pi^{\underline{m}},d\Th^{\underline{\mu}})$ \p{b2},
or as the frame $E^{\underline{A}}$ from Eq. \p{rot}. In the first case
the field strength components
$F_{\underline{AB}}=F_{\underline{AB}}(X,\Th)$
depend on the (super)field $X^{\underline{m}}$ and $\Th^{\underline{\mu}}$
only, while in the second case the field strength components
$F_{\underline{AB}}$ do depend bilinearly on the $u$ and $v$ matrices
\p{a}, \p{aa}
$$
F_{\underline{\a\b}}=v^{\underline{\mu}}_{\underline{\a}}
v^{\underline{\nu}}_{\underline{\mu}}F_{\underline{\mu\nu}}(X,\Th)\equiv
v_{\underline{\a}}Fv_{\underline{\b}}
$$
$$
F_{\underline{\a a}}=v^{\underline{\mu}}_{\underline{\a}}u_{\underline{a}}
^{\underline{m}}F_{\underline{\mu m}}=v_{\underline{\a}}Fu_{\underline{a}}
$$
\begin{equation}\label{baca1}
F_{\underline{ab}}=u_{\underline{a}}^{\underline{m}}u_{\underline{b}}
^{\underline{n}}F_{\underline{mn}}=u_{\underline{a}}Fu_{\underline{b}}
\end{equation}

\subsection{Restrictions on background field.}

Let us begin from a consideration of the complete set of equations of
motion following from the action \p{bac1} with an arbitrary Abelian one --
form
\begin{equation}\label{e1}
{\d S\over \d\rho^{++}}=0\Longrightarrow E^{--}(d)=0
\end{equation}
\begin{equation}\label{e2}
{\d S\over \Om^{--i}(\d)}\equiv u_{\underline{m}}^{i}
{\d S\over \d u_{\underline{m}}^{--}}=0\Longrightarrow \rho^{++}E^{i}(d)=0
\end{equation}
$$
{\d S\over E^{-\dot q}(\d)}\equiv v^{+\underline{\mu}}_{\dot q}
{\d S\over \d\Th^{\underline{\mu}}}=0\Longrightarrow
$$
\begin{equation}\label{e3}
E^{-\dot p}(v^{+}_{\dot p}F
v^{+}_{\dot q}-4i\d_{\dot p\dot q}\rho^{++})
+E^{+q}(v^{-}_{q}Fv^{+}_{\dot
q})+E^{++}({1\over 2}u^{--}Fv^{+}_{\dot q})=0
\end{equation}
$$
{\d S\over E^{--}(\d)}\equiv
u^{\underline{m}++}{\d S\over \d X^{\underline{m}}}=0\Longrightarrow
$$
$$
D\rho^{++}={1\over
4}E^{++}(d)u^{++}Fu^{--}+{1\over 2}E^{+q}(d)u^{++}Fv^{-}_{q}+
{1\over 2}E^{-\dot q}(d)u^{++}Fv^{+}_{q}
$$
$$
{\d S\over E^{i}(\d)}\equiv
u^{\underline{m}i}{\d S\over \d X^{\underline{m}}}=0\Longrightarrow
$$
\begin{equation}\label{e4}
\rho^{++}\Om^{--i}(d)=-{1\over 2}E^{++}(d)u^{i}Fu^{--}
+E^{+q}(d)u^{i}Fv^{-}_{q}+E^{-\dot q}u^{i}Fv^{+}_{\dot q}
\end{equation}
\begin{equation}\label{e5}
{\d S\over E^{++}(\d)}\equiv
u^{\underline{m}--}{\d S\over \d X^{\underline{m}}}=0\Longrightarrow
u^{--}Fv^{-}_{q}E^{+q}(d)+u^{--}Fv^{+}_{\dot q}E^{-\dot q}=0
\end{equation}
\begin{equation}\label{e6}
{\d S\over E^{+q}(\d)}\equiv
v^{-\underline{\mu}}{\d S\over \d \Th^{\underline{\mu}}}=0\Longrightarrow
v^{-}_{q}Fv^{-}_{p}E^{+p}(d)+v^{-}_{q}Fv^{+}_{\dot q}E^{-\dot q}=0
\end{equation}
where the later equations are written with taking into account Eqs.
\p{e1}, \p{e2}.

Equation
\begin{equation}\label{e7}
{\d S\over \Om^{(0)}(\d)}\equiv -u^{--}_{\underline{m}}{\d S\over
\d u^{--}_{\underline{m}}}=0
\end{equation}
is satisfied identically after Eq. \p{e1} is taken into account. This is
Noether identity corresponding to $SO(1,1)$ gauge symmetry of the
generalized action. The Noether identities reflecting $SO(D-2)$ and
$K_{D-2}$ (boost) symmetry of the generalized action are
\begin{equation}\label{e8}
{\d S\over \Om^{ij}(\d)}\equiv 0\Longleftrightarrow \ 0=0
\end{equation}
\begin{equation}\label{e9}
{\d S\over \Om^{++i}(\d)}\equiv 0\Longleftrightarrow \ 0=0
\end{equation}

As in the free superparticle case, they appear due to the absence of
moving frame vectors $u^{++}_{\underline{m}}$ and $u^{i}_{\underline{m}}$
in the generalized action \p{bac1}.

Eq. \p{e6} and \p{e5} are satisfied identically in the absence of
background field. These are the Noether identities reflecting worldline
supersymmetry (realized nonlinearly in the form of irreducible superfield
$\k$ -- symmetry) and reparametrization symmetry (realized as $b$ --
symmetry) of the generalized action \p{c1}.

The supposition about continuous zero field limit requires the existence
of such symmetries for nonvanishing background (see Ref. \cite{shapiro}
and Refs. therein). Following these conjecture we shall require the
background field to be restricted by constraints which shall turn
Eqs. \p{e6}, \p{e5} into identities after Eqs. \p{e1} -- \p{e4} are taken
into account. So Eqs. \p{e6}, \p{e5} shall be treated as equations for
background, which shall be satisfied for any values of worldline
superfields, restricted by Eqs. \p{e1} -- \p{e4}.

Because the Abelian connection form coefficients \p{baca} are supposed to
be independent on harmonic superfields $u$ and $v$, the general solution
of Eq. \p{e6} with respect to the field strength is
\begin{equation}\label{e10}
F_{\underline{\mu\nu}}(X,\Th)=0
\end{equation}

Indeed, this relation results in
\begin{equation}\label{e11}
v^{-}_{q}Fv^{-}_{p}=0
\end{equation}
as well as in
\begin{equation}\label{e12}
v^{-}_{q}Fv^{+}_{\dot q}=0;\ \ \ \ \ v^{+}_{\dot q}Fv^{+}_{\dot p}=0
\end{equation}

To prove that \p{e10} leads also to vanishing of $u^{--}Fv^{-}_{q}$ we
shall take into account the nontrivial parts of the Bianchi
identities for the Abelian field strength
\begin{equation}\label{e13}
dF=0
\end{equation}
which acquire the form
\begin{equation}\label{e14}
\G^{\underline{m}}_{\{\underline{\nu}
\underline{\rho}\vert}F_{\underline{m}\vert\underline{\mu}\}}=0
\end{equation}
\begin{equation}\label{e15}
\bigtriangledown_{\{\underline{\mu}}
F_{\underline{\nu}\}\underline{n}}-2i\G^{\underline{m}}_
{\underline{\mu\nu}}F_{\underline{mn}}=0
\end{equation}
with $\bigtriangledown_{\underline{\mu}}$ being a target superspace
covariant derivative.

Contracting Eq. \p{e14} with harmonics $v^{-\underline{\mu}}_{q}
v^{-\underline{\nu}}_{p}v^{-\underline{\rho}}_{r}$ we get
\begin{equation}\label{e16}
0=v^{-\underline{\mu}}_{\{q}v^{-\underline{\nu}}_{p}
v^{-\underline{\rho}}_{r\}}\G^{\underline{m}}_{\underline{\mu\nu}}
F_{\underline{\rho m}}
=\d_{\{qp}v^{-}_{r\}}Fu^{--}\Longrightarrow v^{-}_{q}Fu^{--}=0
\end{equation}

For the further consideration the following results of
investigation of the Bianchi identities \p{e14}, \p{e15} will be useful.
The general solution of Eq.
\p{e14} is (see \cite{shapiro} and Refs. therein)
\begin{equation}\label{e17}
F_{\underline{m}\underline{\mu}}=(\G_{\underline{m}})_
{\underline{\mu}\underline{\nu}}W^{\underline{\nu}}
\end{equation}
with a Grassmann superfield $W^{\underline{\nu}}$. Then Eq. \p{e15}
acquires the form
\begin{equation}\label{e18}
(\G_{\underline{m}})_{\{\underline{\mu}\vert
\underline{\rho}}\bigtriangledown_{\vert \underline{\nu}\}}
W^{\underline{\rho}}-i\G^{\underline{n}}_{\underline{\mu\nu}}
F_{\underline{nm}}=0
\end{equation}

In the most complicated D=10 case, Eq. \p{e18} expresses the
bosonic component $F_{\underline{mn}}$ in terms of spinor target --
space derivative of superfield $W$
\begin{equation}\label{e19}
F_{\underline{mn}}={i\over 2(D-2)}\bigtriangledown_{\underline{\mu}}
W^{\underline{\nu}}(\G_{\underline{mn}})_{\underline{\nu}}^{\underline{\mu}}
\end{equation}
and, besides, puts the following restrictions on $W$ superfield
$$
\bigtriangledown_{\underline{\mu}}W^{\underline{\mu}}=0
$$
\begin{equation}\label{e20}
\bigtriangledown_{\underline{\mu}}W^{\underline{\nu}}
(\G_{\underline{m}_{1}\dots \underline{m}_{4}})_{\underline{\nu}}^
{\underline{\mu}}=0
\end{equation}

As a result of \p{e20}, Eq. \p{e19} can be solved with respect to
$\bigtriangledown_{\underline{\mu}}W^{\underline{\nu}}$ by
\begin{equation}\label{e21}
\bigtriangledown_{\underline{\mu}}W^{\underline{\nu}}={i\over 2}
F^{\underline{mn}}(\G_{\underline{mn}})_{\underline{\mu}}^{\underline{\nu}}
\end{equation}

To justify that Eq. \p{e5} with background, defined by the constraint
\p{e10}
\begin{equation}\label{e22}
u^{--}Fv^{+}_{\dot q}E^{-\dot q}=0,
\end{equation}
is also satisfied identically, Eq. \p{e3} with such background
\begin{equation}\label{e23}
4i\rho^{++}E^{-\dot q}(d)=-{1\over 2}v^{+}_{\dot q}Fu^{--}E^{++}(d)
\end{equation}
shall be used
$$
u^{--}Fv^{+}_{\dot q}E^{-\dot q}(d)=-{i\over 8\rho^{++}}u^{--}Fv^{+}_{\dot
q} u^{--}Fv^{+}_{\dot q}E^{++}(d)\equiv 0
$$

Hence, for the background satisfying the standard constraint \p{e10}, all
the symmetries of free superparticle action are conserved.

Indeed, the variation of the action \p{bac1} involving backgroung field
\p{baca} satisfying \p{e10} acquires the form
$$
\d S\vert_{F_{\underline{\mu\nu}}=0}=\int_{{\cal{M}}^1}\,\,(\d\rho^{++}-
\rho^{++}\Om^{(0)}(\d)+E^{++}(\d)u^{--}Fu^{++}
$$
$$
-E^{i}(\d)u^{i}Fu^{++}+E^{+q}(\d)v^{-}_{q}Fu^{++})E^{--}(d)
$$
$$
+\int_{{\cal{M}}^1}\,\,(\rho^{++}\Om^{--i}(\d)-E^{\underline{a}}(\d)
u_{\underline{a}}Fu^{i}-E^{\underline{\a}}(\d)v_{\underline{\a}}Fu^{i})
E^{i}(d)
$$
$$
+\int_{{\cal{M}}^1}\,\,(E^{-\dot q}(\d)+{i\over \rho^{++}}E^{++}(\d)u^{--}
Fv^{+}_{\dot q})(-4i\rho^{++}E^{-\dot q}+v^{+}_{\dot q}Fu^{--}E^{++}(d))
$$
$$
-\int_{{\cal{M}}^1}\,\,E^{i}(\d)(\rho^{++}\Om^{--i}(d)+{1\over
2}u^{i}Fu^{--} E^{++}(d)-u^{i}Fv^{-}_{q}E^{+q}-u^{i}Fv^{+}_{\dot
q}E^{-\dot q})
$$
\begin{equation}\label{e24}
-\int_{{\cal{M}}^1}\,\,E^{--}(\d)(D\rho^{++}-{1\over 4}u^{++}Fu^{--}
E^{++}(d)+{1\over 2}u^{++}Fv^{-}_{q}E^{--}(d))
\end{equation}
and contains essentially only the same independent variations
$\d\rho^{++}$, $\Om^{--i}(\d)$, $E^{-\dot q}(\d)$, $E^{--}(\d)$ as a
variation of free superparticle action \p{c4} does (for brevity the
explicit expressions for the field strength components in terms of
$W^{\underline{\mu}}$ superfield are not specified in \p{e24}).

\subsection{Extrinsic geometry of N=1 superparticle in Abelian
background.}

Equations of motion derived from \p{e24} by a variation over
$\rho^{++}$, $\Om^{--i}(\d)$, $E^{--i}(\d)$, $E^{i}(\d)$ and $E^{-\dot
q}(\d)=\d\Th v^{-}_{\dot q}$ variables have the following form:
\begin{equation}\label{bac7}
E^{--}=0;\ \ \ \ \ E^{i}=0
\end{equation}
\begin{equation}\label{bac20}
E^{-\dot q}={1\over 8i\rho^{++}}E^{++}u^{--}Fv^{+}_{\dot q}=
{1\over 4i\rho^{++}}e^{++}W^{\underline{\nu}}v^{-}_{\underline{\nu}\dot
q}
\end{equation}
$$
\Om^{--i}(d)={1\over \rho^{++}}E^{\underline{B}}
F_{\underline{Bm}}u^{i\underline{m}}
$$
\begin{equation}\label{bac9}
={1\over \rho^{++}}e^{+q}\g^{i}_{q\dot q}Wv^{-}_{\dot q}+{i\over
16}e^{++}({1\over \rho^{++}}W^{-}_{q}\bigtriangledown Wv^{+}_{\dot
q}\g^{i}_{q\dot q}+{4\over (\rho^{++})^{2}}Wv^{+}_{q}\g^{i}_{q\dot
q}Wv^{-}_{\dot q})
\end{equation}
\begin{equation}\label{bac8}
D\rho^{++}=-{1\over 2}E^{\underline{B}}
F_{\underline{Bm}}u^{++\underline{m}}=
e^{+q}W^{\underline{\mu}}v^{+}_{\underline{\mu}q}-{i\over 2(D-2)}e^{++}
v^{+\underline{\nu}}_{\dot q}\bigtriangledown_{\underline{\nu}}
W^{\underline{\mu}}v^{-}_{\underline{\mu}\dot q}
\end{equation}
where the conventional rheotropic conditions
$$
\Om^{(0)}(D)=0;\ \ \ \ \ \Om^{ij}(D)=0
$$
\begin{equation}\label{g1}
E^{++}(d)=e^{++}(d);\ \ \ \ \ E^{+q}(d)=e^{+q}(d)
\end{equation}
are chosen in the same form as for the free superparticle case (see
subsection 2.2).

It is easy to see that for superparticle interacting with Abelian gauge
background the master equations \p{c15}, \p{c23}
\begin{equation}\label{g2}
\Pi^{\underline{m}}={1\over 2}e^{++}u^{--\underline{m}}
\end{equation}
\begin{equation}\label{g3}
d\Th^{\underline{\mu}}=e^{+q}v^{-{\underline{\mu}}}_{q}+e^{++}\psi_{++
\dot{q}}^{-}v^{+{\underline{\mu}}}_{\dot q}
\end{equation}
describing superparticle theory off mass -- shell, are satisfied.

Eqs. \p{bac20}, \p{g3} identify $\psi^{-}_{++\dot q}$ superfield with
the main background field strength $W^{\underline{\mu}}$ contracted with
the harmonic $v^{-}_{\underline{\mu}\dot q}$
\begin{equation}\label{g4}
\psi^{-}_{++\dot q}={1\over 4i\rho^{++}}W^{\underline{\mu}}
v^{-}_{\underline{\mu}\dot q}
\end{equation}
Taking it into account we can use Eqs. \p{aa1} -- \p{aa8} for the
description of superparticle motion in Abelian background.

The only problem which can be pointed by the careful reader is related
to the Eq. \p{aa3} which, after substitution \p{g4} acquire the form
\begin{equation}\label{g5}
D_{+q}({1\over \rho^{++}}W^{\underline{\mu}}v^{-}_{\underline{\mu}\dot q})
=\g^{i}_{q\dot q}{1\over D-2}\g^{i}_{p\dot p}
D_{+p}({1\over \rho^{++}}W^{\underline{\mu}}v^{-}_{\underline{\mu}\dot p})
\end{equation}
and seems to be able to produce an additional restriction on the
background superfield $W^{\underline{\mu}}(X,\Th)$. However, it can be
proved (see Appendix) that Eq. \p{g5} is satisfied identically, when Eqs.
\p{bac8}, \p{e21}, \p{a222}, \p{a22}, \p{g4} are taken into account.

Hence, superparticle motion in the Abelian Yang -- Mills background,
satisfying the standard constraints \p{e10}, is described by
\begin{enumerate}
\item
Worldline supergravity $(e^{++},e^{+q})$ satisfying the torsion constraint
$$
T^{++}=-2ie^{+q}e^{+q};\ \ \ \ \
T^{+q}={1\over 8i\rho^{++}}e^{++}\Om^{++i}\g^{i}_{q\dot
q}W^{\underline{\mu}}v^{-}_{\underline{\mu}\dot q}
$$
\item
Covariant one -- form
$$
\Om^{--i}=
{1\over \rho^{++}}e^{+q}\g^{i}_{q\dot q}Wv^{-}_{\dot q}+{i\over
16}e^{++}({1\over \rho^{++}}W^{-}_{q}\bigtriangledown Wv^{+}_{\dot
q}\g^{i}_{q\dot q}+{4\over (\rho^{++})^{2}}Wv^{+}_{q}\g^{i}_{q\dot
q}Wv^{-}_{\dot q})
$$
\item
Covariant one -- form $\Om^{++i}$ \p{aa1}
$$
\Om^{++i}=(e^{+q}-{i\over 2(D-2)}e^{++}D_{+q})\Om^{++i}_{+q}
$$
which is restricted only by the Peterson -- Codazzi equations
$$
D\Om^{++i}=0\Longrightarrow\ D_{+\{q}\Om_{+p\}}=2i \d_{qp}\Om^{++i}_
{++}
$$
and is defined up to the boost symmetry transformations \p{c91}
$$
\d\Om^{++i}_{+p}=D_{+p}k^{++i}
$$
\item
$SO(1,1)$ (spin) connection one -- form $\Om^{(0)}$ expressed in terms of
$W^{\underline{\mu}}$, harmonics and $SO(1,1)$ compensator $\rho^{++}$ by
Eq. \p{bac8}
$$
\Om^{(0)}(d)={1\over \rho^{++}}(d\rho^{++}+e^{+q}Wv^{+}_{q}+{i\over
2(D-2)}e^{++}v^{-}_{q}\bigtriangledown Wv^{+}_{q})
$$
$SO(1,1)$ curvature of these forms is determined by the Gauss equation
$$
{\cal F} \equiv d \Om^{(0)} = {1\over 2} \Om^{--i} \Om^{++i}
$$
with the forms $\Om^{\pm\pm i}$ specified above
\item
$SO(D-2)$ connection one -- form $\Om^{ij}$ restricted by the Ricci
equation \footnote{In the case of extended worldline supersymmetry the
higher components (of dimension $>+3/2$) of these equations are dependent
that can be seen by studying "identities for identities" being the
integrability conditions for the Peterson -- Codazzi, Gauss and Ricci
equations (see ref. \cite{zima} and Refs. therein). For the case of n=1
worldline supersymmetry, realizing in the D=3 superparticle theory, one
irreducible part of the dimension $+2$ identity is independent (see
Appendix C of ref. \cite{bpstv}).}
$$
R^{ij}  \equiv d \Om^{ij} + \Om^{ik} \Om^{kj} =
- \Om^{--[i} \Om^{++j]}
$$
\end{enumerate}

The only fact, which should be stressed, is that all
equations describing the superparticle motion in Abelian background can be
regarded as ones following from the Peterson -- Codazzi, Gauss and Ricci
equations for the forms $\Om^{--i}$, $\Om^{++i}$, $\Om^{(0)}$ and
$\Om^{ij}$ specified above. Furthermore, the latter equations can be
collected into one zero curvature representation, which is given by the
Maurer -- Cartan equations \p{b21}. This reflects the fact that
(super)particle equations in external field can be always solved, at
least formally.

\section{Conclusion.}

In conclusion, we have constructed the generalized action principle for
N=1 superparticle in flat target superspace of bosonic dimensions D=3,4,6
and 10. We have developed the general framework of the doubly
supersymmetric geometrical approach for superparticle and have
investigated the general aspects of superparticle interaction with Abelian
Yang -- Mills background. It was shown that the interaction with
superparticle put the restrictions on the Abelian background superfield
being the standard constraints of supersymmetric gauge theory (see
\cite{shapiro} and Refs. therein).

We have demonstrated that transition to analytical basis can be done in
such a way that the dependence on the coordinates $\Th^{+q}$ and $X^{++}$
(having the properties of compensators with respect to irreducible
$\k$ -- symmetry and $b$ -- symmetry) disappears from the generalized
action.

The presented results can be regarded as a basis for the future attempts
to find a way for covariant quantization of superparticles and
superstrings in the frame of the generalized action concept, as well as a
toy model for studying a general feature of geometrical approach for
supersymmetric objects interacting with natural (supergravity and super --
Yang -- Mills) background.

The most straightforward development of the results of this work consists
in the construction of the generalized action and geometrical approach for
N=2 superparticles \footnote{We shall stress that the action for N=IIA,
IIB D=10 superparticle does not exist in the standard (STVZ -- like)
doubly supersymmetric approach.} as well as for N=1 null -- super -- p --
branes (tensionless super -- p -- branes). These problems are under
investigations now.

At the same time, N=IIA superparticle is the simplest member of N=2
supersymmetric solitons family in D=10 \cite{duff}, for higher members of
which any actions are unknown up to now.

\vspace{1cm}
{\Large\bf Acknowledgments.}

The early stages of this work was discussed with \fbox{Dmitrij Volkov}
whose activity initiated this direction of research. With acknowledgments
to our Teacher we devote this paper to his blessed memory.
Authors are thankful to Dmitri Sorokin, Anatoly Pashnev and Mario Tonin
for fruitful discussions.

I.B. thanks INTAS and Dutch Government for the support of this work
under the Grant N 94 -- 2317.

A.N. thanks Prof. G. Veneziano for the kind hospitality
at the International Conference under the INTAS Grant N 93 -- 493, Torino,
May 2 -- 3, 1996 and Prof. M. Virasoro for the kind hospitality at the 
International Centre for Theoretical Physics.

\vspace{1cm}
{\Large\bf Appendix}

Here we will prove that the restriction
\begin{equation}\label{ap1}
D_{+q}{\psi_{++\dot q}}^{-}=\g^{i}_{q\dot q}{1\over D-2}\g^{i}_{p\dot p}
D_{+p}{\psi_{++\dot p}}^{-}
\end{equation}
is satisfied identically for
\begin{equation}\label{ap2}
{\psi_{++\dot q}}^{-}=-{i\over 4\rho^{++}}W^{\underline{\mu}}
v^{-}_{\underline{\mu}\dot q}
\end{equation}
where $W^{\underline{\mu}}$ is the Abelian Yang -- Mills field strength
\p{e17}, satisfying
\begin{equation}\label{ap3}
\bigtriangledown_{\underline{\mu}}W^{\underline{\mu}}=0
\end{equation}
\begin{equation}\label{ap4}
\bigtriangledown_{\underline{\mu}}W^{\underline{\nu}}={i\over 2}
F^{\underline{mn}}(\G_{\underline{mn}})_{\underline{\mu}}^{\underline{\nu}}
\end{equation}
due to Bianchi identities and Eq. \p{e10}.
Eqs. \p{ap3}, \p{ap4} involve, in particular, equation of motion.

To prove Eq. \p{ap1} for superfield \p{ap2} in D=10 case, it is enough to
prove that
\begin{equation}\label{124}
4i(\g^{j_{1}j_{2}j_{3}})_{q\dot q}D_{+q}{\psi_{++\dot q}}^{-}\equiv
(\g^{j_{1}j_{2}j_{3}})_{q\dot q}
D_{+q}({1\over \rho^{++}}W^{\underline{\mu}}v^{-}_{\underline{\mu}\dot q})
=0
\end{equation}

Taking into account Eqs. \p{bac9}, \p{bac8}, \p{aa5}, \p{aa6}, \p{a22},
\p{c22}, \p{emb}, \p{a18} we can write
$$
4iD_{+q}{\psi_{++\dot q}}^{-}=
D_{+q}({1\over \rho^{++}}W^{\underline{\mu}}v^{-}_{\underline{\mu}\dot q})
= -{1\over (\rho^{++})^{2}}D_{+q}\rho^{++}
W^{\underline{\mu}}v^{-}_{\underline{\mu}\dot q}
$$
$$
+{1\over \rho^{++}}D_{+q}{\cal Z}^{\underline{M}}\bigtriangledown_
{\underline{M}}W^{\underline{\mu}}v^{-}_{\underline{\mu}\dot q}-
{1\over \rho^{++}}W^{\underline{\mu}}D_{+q}v^{-}_{\underline{\mu}\dot q}
$$
$$
={2\over (\rho^{++})^{2}}W^{\underline{\mu}}v^{+}_{\underline{\mu} q}
W^{\underline{\mu}}v^{-}_{\underline{\mu}\dot q}-{1\over (\rho^{++})^{2}}
\g^{i}_{q\dot q}\g^{i}_{p\dot p}W^{\underline{\mu}}v^{+}_{\underline{\mu}p}
W^{\underline{\mu}}v^{-}_{\underline{\mu}\dot q}
$$
\begin{equation}\label{123}
+{1\over (\rho^{++})^{2}}v^{-\underline{\mu}}_{q}
v^{-}_{\underline{\nu}\dot q}\bigtriangledown_
{\underline{\mu}}W^{\underline{\nu}}
\end{equation}

Contracting \p{123} with $(\g^{klm})_{q\dot q}$, one could find, that the
contributions from the first two terms in r.h.s. of Eq. \p{123} into Eq.
\p{124} cancel one another due to identity
$$
({\tilde\g}^{i}\g^{klm}{\tilde\g}^{i})_{q\dot q}=2(\g^{klm})_{q\dot q}
$$
which holds for the eight -- dimensional $\sigma$ -- matrices
$\g^{i}_{q\dot q}$.

The contribution from the third term becomes proportional to
$$
(\g^{ij^\prime j})_{q\dot q}v^{-\underline{\mu}}_{q}(\G_{\underline{mn}})_
{\underline{\mu}}^{\underline{\nu}}v^{-}_{\underline{\nu}\dot q}
$$
when Eq. \p{ap4} is taken into account.
To prove that it vanishes, let us perform the following algebraic
manipulations using the condition of $\G$ -- matrices invariance \p{b9}
$$
v^{\underline{\mu}}_{\underline{\a}}(\G_{\underline{m}}\G_{\underline{n}})
_{\underline{\mu}}^{\underline{\nu}}v^{\underline{\b}}_{\underline{\nu}}
=\d_{\underline{\a}}^{\underline{\b}}+v^{\underline{\mu}}_{\underline{\a}}
(\G_{\underline{mn}})
_{\underline{\mu}}^{\underline{\nu}}v^{\underline{\b}}_{\underline{\nu}}
=v^{\underline{\mu}}_{\underline{\a}}(\G_{\underline{m}})_
{\underline{\mu\nu}}v_{\underline{\g}}^{\underline{\rho}}
v^{\underline{\g}}_{\underline{\rho}}(\G_{\underline{n}})^
{\underline{\mu\nu}}v^{\underline{\b}}_{\underline{\nu}}
$$
\begin{equation}\label{*}
=u^{\underline{a}}_{\underline{m}}(\G_{\underline{m}})_
{\underline{\a\g}}(\G_{\underline{b}})^{\underline{\g\b}}
u^{\underline{b}}_{\underline{n}}=u^{\underline{a}}_{\underline{m}}
u^{\underline{b}}_{\underline{n}}(\G_{\underline{a}}\G_{\underline{b}})
_{\underline{\a}}^{\underline{\b}}
\end{equation}

Taking in Eq. \p{*} ${\underline{\a}}={(^{-}_{q}})$,
${\underline{\b}}=(-\dot q)$ and using the explicit representation for
$\G$ -- matrices (see \cite{bzst,bzp}), we get
$$
v^{-\underline{\mu}}_{q}(\G_{\underline{mn}})_{\underline{\mu}}^
{\underline{\nu}}v^{-}_{\underline{\nu}\dot q}=2u^{--}_{[\underline{m}}
u^{i}_{\underline{n}]}\g^{i}_{q\dot q}
$$
and, hence
$$
(\g^{j_{1}j_{2}j_{3}})_{q\dot q}
v^{-\underline{\mu}}_{q}(\G_{\underline{mn}})_
{\underline{\mu}}^{\underline{\nu}}v^{-}_{\underline{\nu}\dot q}=
2u^{--}_{[\underline{m}}u^{i}_{\underline{n}]}\ Tr\ ({\tilde \g}^{i}
\g^{j_{1}j_{2}j_{3}})=0
$$

So, we have proved that superfield \p{ap2} satisfies the relation
\p{ap1}.


\begin{thebibliography}{99}

\bibitem{bsv}
I. Bandos, D. Sorokin and D. Volkov,
{\sl Phys.Lett.}{\bf B352} (1995) 269
{\bf (hep-th/9502141)};\\
D. Volkov, The generalized action principle for superstrings and
supermembranes, {\sl Proceedings of the International Conference
"SUSY -- 95", Paris, France, May 1995}, {\bf hep-th/9512103};\\
I.Bandos, Doubly supersymmetric geometric approach for heterotic string:
from Generalized Action Principle to exactly solvable nonlinear equations,
{\sl Proceedings of the International Conference, Rahov, Ukraine,
September 1995}, {\bf hep-th/9510213},\\
I. Bandos, Generalized action principle and geometric approach for
superstrings and super -- p -- branes, {\sl Proceedings of the
International Conference "Alushta -- 96", Alushta, Ukraine, May 1996},
{\bf hep-th/9608094};\\
A. Nurmagambetov, A generalized action principle for D=4 doubly
supersymmetric membranes, {\sl Proceedings of Xth International Workshop
 on HEP\&QFT, September 19--25 1995, Zvenigorod, Russia}, {\bf
 hep-th/9512102}.

\bibitem{stv}
D. Sorokin, V. Tkach and D. V. Volkov, {\sl Mod. Phys. Lett.} {\bf A4}
(1989) 901.

\bibitem{vz}
D. V. Volkov and A. Zheltukhin, \JETPL 48 1988 61;
\LMP 17 1989 141; \NPB 335 1990 723.

\bibitem{stvz}
D. Sorokin, V. Tkach, D. V. Volkov and A. Zheltukhin, {\sl Phys. Lett.}
{\bf B216} (1989) 302.

\bibitem{sp}
F. Delduc and E. Sokatchev, {\sl Class. Quantum Grav.} {\bf 9} (1992) 361.

\bibitem{gs92}
A. Galperin and E. Sokatchev, {\sl Phys. Rev.} {\bf D46} (1992) 714.

\bibitem{pashnev}
A.Pashnev and D.Sorokin, \CQG 10 1993 625.

\bibitem{hsstr}
N. Berkovits, {\sl Phys. Lett.} {\bf
B232} (1989) 184; {\bf B241} (1990) 497; {\sl Nucl. Phys.} {\bf B350}
(1991) 193; {\bf B358} (1991) 169; {\sl Nucl. Phys.} {\bf B379} (1992) 96;
{\bf B395} (1993) 77.

\bibitem{iv@kap}
E. Ivanov and A. Kapustnikov {\sl Phys. Lett.}, {\bf B267} (1991) 175.

\bibitem{ivanov}
E.I.Ivanov and A.A.Kapustnikov, {\sl Int.J.Mod.Phys.} {\bf A7} (1992)
2153.

\bibitem{to}
M. Tonin, {\sl Phys. Lett.} {\bf B266} (1991) 312; \\
{\sl Int. J. Mod.Phys} {\bf 7} 1992 613; \\
S. Aoyama, P. Pasti and M. Tonin, {\sl Phys. Lett.} {\bf
B283} (1992) 213.

\bibitem{dis}
F. Delduc, E. Ivanov and E. Sokatchev, \NPB 384 1992 334.

\bibitem{dghs92}
F. Delduc, A. Galperin, P. Howe and E. Sokatchev, {\sl Phys. Rev.}
{\bf D47} (1992) 587.

\bibitem{gs93}
V. Chikalov and A. Pashnev, {\sl Mod.  Phys.  Lett.} {\bf
A8} (1993) 285.

\bibitem{gs2}
A. Galperin and E. Sokatchev, \PRD 48 1993 4810.

\bibitem{tp93}
P. Pasti and M. Tonin, \NPB 418 1994 337.

\bibitem{bers94}
E. Bergshoeff and E. Sezgin, \NPB 422 1994 329.\\
E. Sezgin, Preprint CTP TAMU--58/94, November 1994; {\bf hep--th/9411055}.

\bibitem{bnsv}
I.Bandos, A.Nurmagambetov, D.Sorokin and D.Volkov,
{\sl JETP Lett.} {\bf 60} (1994) 613, {\bf hep-th/9409439};\\
I.Bandos, A.Nurmagambetov, D.Sorokin and D.Volkov,
{\sl Class.Quantum Grav.} {\bf 12} (1995) 1881, {\bf hep-th/9502143}.

\bibitem{bpstv}
I. Bandos, P. Pasti, D. Sorokin, M. Tonin and D. Volkov,
{\sl Nucl.Phys.} {\bf B446} (1995) 79, {\bf hep-th/9501113}.

\bibitem{berk}
N.Berkovits, \NPB 431 1994 258.

\bibitem{nissimov}
E.Nissimov, S.Pacheva and S.Solomon, \NPB 296 1988 462.

\bibitem{kallosh}
R.Kallosh and M.Rahmanov, {\sl Phys.Lett.} {\bf B209} (1988) 233;
{\sl Phys.Lett.} {\bf B214} (1988) 549;\
R.Kallosh, {\sl Phys.Lett.} {\bf 225} (1989) 49.

\bibitem{bzst}
I. A. Bandos and A. A. Zheltukhin,
 {\sl JETP. Lett.} {\bf 54} (1991) 421;
 {\sl Phys. Lett.} {\bf B288} (1992) 77;
 {\sl Phys. Atom. Nucl.} {\bf 56} (1993)
 113--121;
{\sl Phys. Part. Nucl.} {\bf 25} (1994) 453--477.

\bibitem{bhbz0}
I. A. Bandos, {\sl Sov. J. Nucl. Phys.} {\bf 51} (1990) 906;
 {\sl JETP. Lett.} {\bf 52} (1990) 205.\\
I. A. Bandos and A. A. Zheltukhin , {\sl JETP. Lett.} {\bf 51} (1990)
547; {\bf 53} (1991) 7; {\sl Phys. Lett.} {\bf B261}
 (1991) 245;
 {\sl Theor. Math. Phys.} {\bf 88} (1991) 358;
 {\sl Fortschr. Phys.} {\bf 41} (1993) 619--676.

\bibitem{bzm}
I. A. Bandos and A. A. Zheltukhin,
 {\sl JETP. Lett.} {\bf 55} (1992) 81;
 {\sl Int. J. Mod. Phys.} {\bf A8} (1993) 1081-- 1091.

\bibitem{bzp}
I. A. Bandos and A. A. Zheltukhin ,
{\sl Class. Quantum Grav.} {\bf 12} (1995) 609--626
 {\bf (hep-th/9405113)}.

\bibitem{kallosh87}
R.Kallosh, {\sl JETP Lett.} {\bf 45} (1987) 365; \PLB 195 1987 369.

\bibitem{morozov88}
R.Kallosh and A.Morozov, {\sl Int.J.Mod.Phys.} {\bf A3} (1988) 1943.

\bibitem{kallosh95}
R.Kallosh, \PRD 52 1995 6020,
Preprint {\bf SU-ITP-95-12}, {\bf hep-th/9506113}.

\bibitem{carlip}
S.Carlip, \NPB 284 1987 365.

\bibitem{bandzer}
I.Bandos, Preprint {\bf hep-th/9510216} (Submitted to {\sl Phys.Lett.B}).

\bibitem{bsv96}
I.Bandos, D.Sorokin and D.Volkov, \PLB 372 1996 77.

\bibitem{geo}
F. Lund and T. Regge, \PRD 14 1976 1524.\\
R.Omnes \NPB 149 1979 269.\\
B. M. Barabashov and V. V. Nesterenko, {\sl Relativistic string model in
hadron physics}, Moscow, Energoatomizdat, 1987 (and refs. therein).\\
A. Zheltukhin, \SJNP 33 1981 1723 ; \TMP 52 1982 73 ;
\PLB 116 1982 147; \TMP 56 1983 230.

\bibitem{toppan}
D.Sorokin and F.Toppan, Preprint {\bf hep-th/9603187}, {\sl Nucl. Phys}
{\bf 480} (1996) n 1,2 (in press).

\bibitem{bakas}
I.Bakas and K.Sfetsos,
Preprint {\bf CERN-TH/96-89}, {\bf hep-th/9604195}.

\bibitem{viswan}
K.Viswanathan and R.Parthasarathy, Preprint {\bf SFU-HEP-04-96}.

\bibitem{z0}
A. Zheltukhin, {\sl JETP Lett.} {\bf 46} (1987) 208; {\sl
Sov.J.Nucl.Phys.} {\bf 48} (1988) 375.

\bibitem{zhelt}
A.A.Zheltukhin, Preprint {\bf hep-th/9606013} (Submitted to Class. Quantum
Grav.).

\bibitem{sanchez}
C.O.Lousto, N.Sanchez, Preprint {\bf gr-qc/9605015}.

\bibitem{39prime}
O.J.Ganor and A.Hanany, Preprint {\bf hep-th/9602120};\\
N.Seiberg and E.Witten, Preprint {\bf hep-th/9603003};\\
M.J.Duff, H.L{\"u} and C.N.Pope, Preprint {\bf hep-th/9603037}.

\bibitem{rheo}
Y. Nieman and T. Regge, \PLB 74 1978 31, {\sl Revista del Nuovo Cim.}
1 1978 1;
R. \'D Auria, P. Fr\'e and T. Regge, {\sl Revista del Nuovo Cim.}
3 1980 1;\\
L. Castellani, R. \'D Auria, P. Fr\'e. ``Supergravity and superstrings, a
geometric perspective'', World Scientific, Singapore, 1991 (and
references therein).

\bibitem{sok}
E. Sokatchev , {\sl Phys. Lett.} {\bf B169} (1987) 209.

\bibitem{sok87}
E. Sokatchev , \CQG 4 1987 237.

\bibitem{shapiro}
J.A.Shapiro and C.C.Taylor, {\sl Phys.Rep.} {\bf 191} (1990) 221.

\bibitem{k}
J. De Azcarraga and J.Lukiersky, {\sl Phys. Lett.} {\bf B113} (1982)
170.\\
W.Siegel, {\sl Phys. Lett.} {\bf B128} (1983) 397.

\bibitem{volkov}
V. P. Akulov and  D. V. Volkov,  \JETPL 16 1972 438; \PLB 46 1973 109.

\bibitem{newman}
E.Newman, R.Penrose, {\sl J.Math.Phys.} {\bf 3} (1962) 566.

\bibitem{ghs}
A. Galperin, P. Howe and K. Stelle, \NPB 368 1992 248;
A. Galperin, F. Delduc and E. Sokatchev, \NPB 368 1992 143;
A. Galperin, P. Howe and P. Townsend, \NPB 402 1993 531.

\bibitem{rindler}
R.Penrose and W.Rindler, {\sl Spinors and Space -- Time}, v. 1,2,
Cambridge University Press, 1986.

\bibitem{penrose}
R.Penrose and M.McCallum, {\sl Phys.Rep.} {\bf 6} (1972) 241.

\bibitem{zima}
I.A.Bandos and V.G.Zima, {\sl Theor.Math.Phys.} {\bf 70} (1987) 52.

\bibitem{duff}
M.J. Duff, Ramzi R.Khuri and J.X.Lu, {\sl Phys.Rep.} {\bf 259} (1995) 213.

\bibitem{sezgin}
P.S.Howe and E.Sezgin, {\sl Superbranes}, Preprint {\bf CERN-TH/96-200},
{\bf hep-th/9607227}.

\bibitem{derizl}
A.A.Deriglazov, A.V.Galajinsky and S.L.Lyakhovich, Preprint {\bf
hep-th/9512036}.

\end{thebibliography}
\end{document}